\def\DIRvalue{Morrison}
\def\IDvalue{MO}
\def\titlevalue{Gromov--Witten invariants and localization}
\def\authorvalue{David R. Morrison}
\def\shortauthorvalue{\authorvalue}
\def\addressvalue{Departments of Mathematics and Physics\\
University of California, Santa Barbara\\
Santa Barbara, CA 93106 USA\\
  \tt drm@math.ucsb.edu}

\def\abstractvalue{We give a pedagogical review of the computation of 
Gromov--Witten invariants via localization in 2D gauged linear
sigma models.  
We explain the relationship between the two-sphere partition function
of the theory and the K\"ahler potential on the conformal manifold.
We show how the K\"ahler potential can be assembled from classical,
perturbative, and non-perturbative contributions, and explain how
the non-perturbative contributions are related to the Gromov-Witten
invariants of the corresponding Calabi--Yau manifold.
We then explain how localization 
enables efficient calculation of the two-sphere
partition function and, ultimately,  the Gromov--Witten invariants
themselves.
}

\def\preprintvalue{}

\ifx\ifLONG\undefined
\documentclass[12pt]{article}
\newcommand{\chapterauthor}[1]{
\begin{center}
{\bf \normalsize  #1}
\end{center}
}

\newcommand{\chapteraddress}[1]{
\begin{center}
{ \small \it \addressvalue}
\end{center}
}

\newcommand{\chapterabstract}[1]{
\vspace{\baselineskip}
\begin{center}
\textbf{\small Abstract}
\end{center}
#1}

\newcommand{\chapterheader}{

\chapter[\titlevalue{}  (by \shortauthorvalue)]{\titlevalue}
\label{Chapter\IDvalue}
\chapterauthor{\authorvalue}
\chapteraddress{\addressvalue}
\chapterabstract{\abstractvalue}
\tightmtctrue
\minitoc
}

\newcommand{\documentheader}{
\begin{flushright} \small
  \preprintvalue
 \end{flushright}

\begin{center}
{\bf \Large \titlevalue}
\end{center}

\chapterauthor{\authorvalue}
\chapteraddress{\addressvalue}
\chapterabstract{\abstractvalue}

\medskip

This is a contribution to the review volume ``Localization techniques
in quantum field theories'' (eds. V.~Pestun and M.~Zabzine) which
contains 17 Chapters available at \cite{ContributionSummary}

\tableofcontents
}

\newcommand{\ifvolume}[2]{\ifx\ifLONG\undefined#2\else#1\fi}

\newcommand{\documentfinish}{
\ifx\ifLONG\undefined
\bibliographystyle{bibreview} 
\bibliography{\IDvalue,review}  
\end{document}
\else
\addcontentsline{toc}{section}{References}
\providecommand{\href}[2]{#2}\begingroup\raggedright\endgroup

\fi
}

\newcommand{\documentfinishBBL}{
\addcontentsline{toc}{section}{References}
\ifx\ifLONG\undefined
\input{\IDvalue.separate.bbl}
\end{document}
\else
\input{\DIRvalue/\IDvalue.volume.bbl}
\fi
}

\ifx\ifLONG\undefined

\else 

\fi

\usepackage{cite}
\usepackage{amsmath,amsfonts,graphicx,color,hyperref}
\usepackage{fullpage}

\numberwithin{equation}{section}

\begin{document}
\thispagestyle{empty}
\documentheader
\newcommand{\tr}{\operatorname{tr}}
\else \chapterheader 
\fi

\newcommand{\MObstrut}{\rule[-1.3ex]{0ex}{3.6ex}}

\numberwithin{equation}{section}

\section{Introduction}

Many of the early studies of conformal field theories in two dimensions
were motivated by the connection of these theories
to perturbative string theory.  When the string theory is being
compactified on a compact manifold $X$ (typically a Calabi--Yau manifold),
the resulting conformal field theory can be described in terms of 
the nonlinear sigma model with target space $X$.
One of the interesting features of these theories is the phenomenon of
{\em mirror symmetry}\/ \cite{MOGreene:1990ud,MOCLS,MOAspinwall:1990xe}: 
two different Calabi--Yau manifolds $X$
and $Y$ can lead to conformal field theories which are identical save
for a relabeling of the action of the superconformal algebra.

The celebrated paper of Candelas, de la Ossa, Green, and Parkes \cite{MOCDGP}
exploited mirror symmetry to provide a new way to calculate instanton
contributions to the sigma model (now known as ``Gromov--Witten invariants''
\cite{MOGromov,MOtsm}),
appealing to the fact that instanton-corrected correlation functions in one
theory corresponded to correlation functions in the other theory which
receive no quantum corrections.
This powerful method,
eventually formalized as a mathematical ``Mirror Theorem'' \cite{MOcomplete1,MOcomplete2},
only works when the mirror partner of a given Calabi--Yau manifold is known.
Subsequent developments in mathematics 
(cf.~\cite{MOarXiv:1201.6350}) suggest that it should
be possible to determine the Gromov--Witten invariants without recourse
to the mirror, and that has now been achieved in a physics context
as well \cite{MO2-sphere}.  This new physical method for finding Gromov--Witten
invariants is the subject of the present review.

The method is a by-product of a recent theme in the study of supersymmetric
quantum field theories, which formulates a given theory on a sphere or product
of spheres, and evaluates physical quantities such as the partition
function by means of localization.  This theme was pioneered in four dimensions
by Pestun \cite{MOPestun:2007rz}, and has subsequently been extended to a number
of different dimensions and contexts, many of which are covered 
in \ifvolume{this volume}{the collection of reviews \cite{ContributionSummary}}.

For theories in dimension two with $\mathcal{N}=(2,2)$ supersymmetry,
the formulation on the two-sphere and the corresponding
localization computations were carried out in
\cite{MOBenini:2012ui,MODoroud:2012xw}.
The authors of \cite{MO2-sphere} then recognized that there was a connection
between the partition function on the two-sphere and 
 the Zamolodchikov metric on the conformal manifold of the theory, formulating
this as a precise conjecture.  They also showed how the conjecture would
enable the calculation of Gromov--Witten invariants from the
data of the partition function, without needing a mirror Calabi--Yau
manifold.

Compelling arguments in favor of the conjecture were soon given in
\cite{MOGomis:2012wy,MOGerchkovitz:2014gta}. We present here instead a more
recent argument \cite{MOGomis:2015yaa} which explains the
result as arising from an anomaly of the conformal field theory.
We review that argument in Section~\ref{MOsec:KpZ}.

In Section~\ref{MOsec:metrics} we then discuss two kinds of sigma models:  the nonlinear sigma model
with Calabi--Yau target space (including the classical K\"ahler potential
on the conformal manifold), and the ``gauged linear sigma model'' of 
\cite{MOWitten:1993yc},
which is where we shall carry out our localization computations.
Quantum corrections to the K\"ahler potential, including the non-perturbative
corrections associated to Gromov--Witten invariants,  are discussed in
Section~\ref{MOsec:perturb}.

The determination of the two-sphere partition function via localization, 
and the corresponding
method for calculating Gromov--Witten invariants, is reviewed in
Section~\ref{MOsec:S2}.  Finally, in Section~\ref{MOsec:evaluating}
we discuss the evaluation of the partition function via residues
and show how to obtain Gromov--Witten invariants explicitly.
We have collected supplementary material in 
an Appendix (Section~\ref{MOapp:N22}).

We have drawn heavily upon
\cite{MOsumming,MOtowards-duality},\cite{MO2-sphere},\cite{MOnew-methods,MOgamma-class},\cite{MOGomis:2015yaa}
in preparing this review.

\section{K\"ahler potentials and $2$-sphere partition functions}
\label{MOsec:KpZ}

Let us consider the exactly marginal operators for a two-dimensional
conformal field theory.  These are operators $\mathcal{O}_I$ 
having the property that, if added to
the action with coupling constants $\lambda^I$
\begin{equation}
\delta S = \frac1\pi \sum_I \int d^2x\, \lambda^I \mathcal{O}_I(x) \ ,
\end{equation}
they leave the theory conformally invariant.  The coupling constants $\lambda^I$ 
parameterize the {\em conformal manifold}\/ $\mathcal{M}$ of the theory, and the
two-point functions
\begin{equation}
\langle \mathcal{O}_I(x)\mathcal{O}_J(y) \rangle
= \frac{g_{IJ}(\lambda^K)}{(x-y)^4}
\label{MOeq:two-point}
\end{equation}
determine the {\em Zamolodchikov metric}\/ $g_{IJ}$ on $\mathcal{M}$
\cite{MOZamolodchikov:1986gt}.

In momentum space the two-point functions \eqref{MOeq:two-point} take
the form
\begin{equation}
\langle \mathcal{O}_I(p) \mathcal{O}_J(-p)\rangle
\sim p^4 \log \left(\frac{\Lambda^2}{p^2}\right) \ .
\end{equation}
Having a logarithmic behavior with
cutoff $\Lambda$ does not violate scale invariance since
any rescaling of $\Lambda$ can be compensated with a contact term.
However, although they do not spoil conformal invariance, these logarithms
lead to the non-conservation of the dilatation charge in the 
presence of non-vanishing background fields (the original ``conformal anomaly'').
This can be detected by promoting the couplings $\lambda^I$ to
fields \cite{MOSeiberg:1993vc}.
Then the anomaly induces a term in the energy-momentum trace 
of the rough form
\begin{equation}
T^\mu_\mu = g_{IJ}\lambda^I \Box \lambda^J + \cdots \ .
\end{equation}
In this section, we shall discuss a further conformal anomaly under variation of the
2D metric, following \cite{MOGomis:2015yaa}.  This anomaly was first
observed in \cite{MOOsborn:1991gm}, and is again consistent with
scale invariance  due to the possibility of
contact terms  in the two-point function.

We assume as in \cite{MOGomis:2015yaa} that the given conformal field theory can
be regulated in a dif\-feo\-mor\-phism-invariant way, including a metric $\gamma_{\mu\nu}$
on the 2D spacetime as well as spacetime-dependent couplings $\lambda^I$.
The partition function of the theory on this spacetime
then depends on the metric
and couplings, taking the form $Z[\gamma_{\mu\nu};\lambda^I]$.

We consider an infinitesimal Weyl transformation
\begin{equation}
\delta_\sigma \gamma_{\mu\nu} = 2 \gamma_{\mu\nu}\, \delta\sigma
\end{equation}
(where the infinitesimal $\delta\sigma$ has compact support) and ask for the corresponding
variation $\delta_\sigma \log Z$ of the partition function.
A precise form  of the infinitesimal Weyl variation of $\log Z$
is derived in \cite{MOOsborn:1991gm} and takes the form
\begin{equation}
\delta_\sigma \log Z = 
\frac c{24\pi} \int d^2x\, \delta \sigma\, \sqrt\gamma\, R
-\frac1{4\pi}\int d^2x\, \delta\sigma\, \sqrt\gamma \, g_{IJ}\, \gamma^{\mu\nu}
\partial_\mu\lambda^I \partial_\nu\lambda^J \ ,
\label{MOeq:anomaly-functional}
\end{equation}
where $R$ is the Ricci scalar.  The first term is a universal contribution
due to the central charge $c$ of the theory.
It is argued in \cite{MOGomis:2015yaa} that no  anomalies other than
\eqref{MOeq:anomaly-functional}
are possible.

The ``conformal anomaly'' functional \eqref{MOeq:anomaly-functional} describes a sigma
model with target space $\mathcal{M}$, 
and is not the Weyl
variation of any  local counterterm.  It is therefore cohomologically
nontrivial.

There {\em is} an allowed local counterterm of the form 
\begin{equation}
\int d^2x\, \sqrt\gamma\, R \, F(\lambda^I)
\label{MOeq:counterterm}
\end{equation}
whose Weyl variation is
\begin{equation}
\delta_\sigma \int d^2x\, \sqrt\gamma\, R \, F(\lambda^I)
= -2\int d^2x\,\sqrt\gamma\, \Box\, (\delta\sigma) F(\lambda^I) \ .
\label{MOeq:variation}
\end{equation}
Thus, \eqref{MOeq:anomaly-functional} can be shifted by terms of the
form \eqref{MOeq:variation}.

In the case of an $\mathcal{N}=(2,2)$ theory,\footnote{We establish some 
notation and properties for these theories in an Appendix (Section~\ref{MOapp:N22}).}
 exactly marginal
operators can either be chiral or twisted chiral:
\begin{equation}
\delta S = \frac1\pi \int d^2x\, \left( \sum_I \lambda^I \int d^2\theta\, \mathcal{O}_I(x,\theta) + \sum_A \widetilde{\lambda}^A \int d^2\theta\,
\widetilde{\mathcal{O}}_A(x,\theta) + \text{c.c.}\right),
\end{equation}
where $\mathcal{O}_I$ is chiral and $\widetilde{\mathcal{O}}_A$ is
twisted chiral.
The analysis in \cite{MOGomis:2015yaa} assumes that the parameters 
$\lambda^I$ and $\widetilde{\lambda}^A$ can be promoted to 
background chiral and twisted chiral superfields respectively,
and we make the same assumption.

We wish to supersymmetrize the conformal anomaly \eqref{MOeq:anomaly-functional}
and the counterterm \eqref{MOeq:counterterm}.  In order to do so, we place
the theory in curved superspace \cite{MOFestuccia:2011ws}.  The possibilities
for doing so with $\mathcal{N}=(2,2)$ supersymmetry were analyzed in
\cite{MOClosset:2014pda,MOBae:2015eoa}, and amount to a coupling to supergravity.

There are two distinct supergravities to which we could  couple, known as
$U(1)_V$ and $U(1)_A$ \cite{MOHowe:1987ba}; the label indicates whether
the $U(1)$ symmetry preserved in the Poincar\'e supergravity theory
is vector or axial.  From the point of view of the $\mathcal{N}=(2,2)$ SCFT, the
theory has an $R$-symmetry of the form $U(1)_V\times U(1)_A$, and we
can couple either factor (but not both) to a background gauge field.
As in \cite{MOGomis:2015yaa}, we assume\footnote{This issue is discussed 
in detail in \cite{MOClosset:2014pda} building on the general framework 
of \cite{MOFestuccia:2011ws} (see also \cite{MOAdams:2011vw}).}
that the theory can regularized so as to preserve diffeomorphism
invariance, supersymmetry, and either $U(1)_V$ or $U(1)_A$; once this is
done, the other $R$-symmetry cannot be preserved by the regularization scheme.
In particular, our assumptions imply that there are no gravitational
anomalies and that $c_L=c_R$.

Since every two-dimensional metric is conformally flat, the conformal
factor $\sigma$ may be used to specify the metric.  When we supersymmetrize,
the conformal factor becomes part of
a superfield.  In the case of $U(1)_A$ supergravity,
it is the scalar in a chiral superfield $\Sigma$ while in the case of
$U(1)_V$ supergravity it is the scalar in a twisted chiral superfield 
$\widetilde{\Sigma}$.\footnote{If there is any danger of confusion,
we shall indicate the $R$-symmetry in our notation and refer to these
superfields as $\Sigma_R$ or $\widetilde{\Sigma}_R$.}  We will focus on the case of $U(1)_V$ in this
discussion, although a similar discussion holds for $U(1)_A$ (and was
carried out  in \cite{MOGomis:2015yaa}).

The supersymmetrization of the conformal anomaly \eqref{MOeq:anomaly-functional}
is straightforward.  In the regularization which preserves $U(1)_V$ it takes
the form
\begin{equation}
\delta_{\widetilde{\Sigma}} \log Z_V = 
\frac c{24\pi} \int d^2x\, d^4\theta\, 
(\delta\widetilde{\Sigma}+\delta\overline{\widetilde{\Sigma}})
(\widetilde{\Sigma}+\overline{\widetilde{\Sigma}})
-\frac1{4\pi}\int d^2x\, d^4\theta\, \left( \delta\widetilde{\Sigma} \, 
{K}(\lambda,\overline{\lambda},\widetilde{\lambda},
\overline{\widetilde{\lambda}}) + \text{c.c.} \right) \ , 
\label{MOeq:super-anomaly}
\end{equation}
using superconformal gauge.
In fact, the classification of anomalies allows one to conclude \cite{MOGomis:2015yaa} that 
${K}$ is real and
\begin{equation}
K = K_c(\lambda,\overline{\lambda}) - K_{tc}(\widetilde{\lambda},\overline{\widetilde{\lambda}}) \ ,
\end{equation}
so that $\mathcal{M}=\mathcal{M}_c \times \mathcal{M}_{tc}$ is metrically
a product.\footnote{This product structure holds at smooth points;
at certain singular points we may need to take the quotient of the product by a 
finite group \cite{MOBershadsky:1995sp,MOenhanced}.}
  The K\"ahler potential on $\mathcal{M}_c$ is $K_c$
which depends only on the chiral parameters and the K\"ahler potential
on $\mathcal{M}_{tc}$ is $K_{tc}$ which depends only on the twisted
chiral parameters.  We conclude that the conformal anomaly can be written in the form
\begin{equation}
\begin{aligned}
\delta_{\widetilde{\Sigma}} \log Z_V &= 
\frac c{24\pi} \int d^2x\, d^4\theta\, 
(\delta\widetilde{\Sigma}+\delta\overline{\widetilde{\Sigma}})
(\widetilde{\Sigma}+\overline{\widetilde{\Sigma}}) \\
&\qquad
-\frac1{4\pi}\int d^2x\, d^4\theta\,  (\delta\widetilde{\Sigma} 
+ \delta\overline{\widetilde{\Sigma}})\,
(K_c(\lambda,\overline{\lambda})-K_{tc}(\widetilde{\lambda},
\overline{\widetilde{\lambda}})) \ .
\end{aligned}
\label{MOeq:super-anomaly-product}
\end{equation}

We also need a supersymmetric version of the allowed local counterterm.
In the $U(1)_V$ case this takes the form \cite{MOGerchkovitz:2014gta}
\begin{equation}
S_V = 
\frac1{4\pi}\int d^2x\, d^2\theta\, \widetilde{\mathcal{R}} \, F(\widetilde{\lambda}) + \text{c.c.}
=
\frac1{4\pi}\int d^2x\, d^4\theta\, \overline{\widetilde{\Sigma}} \, F(\widetilde{\lambda}) + \text{c.c.}
\label{MOeq:super-counterterm}
\end{equation}
where $\widetilde{\mathcal{R}}=\overline{D}^2\overline{\widetilde{\Sigma}}$ 
is the twisted chiral curvature superfield in superconformal gauge.
The counterterm \eqref{MOeq:super-counterterm} depends only on the
twisted chiral parameters $\widetilde{\lambda}$ and the dependence
is holomorphic.  Under a super-Weyl transformation,
\begin{equation}
\delta_{\widetilde{\Sigma}} S_V = \frac1{4\pi}
 \int d^2x\, d^4\theta\, 
\left(\delta\overline{\widetilde{\Sigma}}\, F(\widetilde{\lambda}) +
\delta\widetilde{\Sigma}\, \overline{F}(\overline{\widetilde{\lambda}})\right) \ .
\end{equation}

The effect of adding a local counterterm of the form \eqref{MOeq:super-counterterm} is to shift the twisted chiral K\"ahler potential
\begin{equation}
K_{tc} \to K_{tc} + F(\widetilde{\lambda}) + 
\overline{F}(\overline{\widetilde{\lambda}}) \ .
\end{equation}
The chiral K\"ahler potential $K_c$ is unchanged by the addition
of counterterms.

The conformal anomaly will affect the partition function whenever
the theory is placed on a
curved manifold with non-trivial topology.  In particular, for
compactification on a two-sphere, both the dependence on the radius
(via the central charge) and the radius-independent part of
the anomaly will be visible in the partition function.  If we 
compactify so as to preserve the $U(1)_A$ symmetry, the partition
function will detect $K_c(\lambda,\overline{\lambda})$ and be
independent of $\widetilde{\lambda}$; on the other hand, if
we compactify so as to preserve the $U(1)_V$ symmetry (as we will
do here), the partition function takes the form
\begin{equation}
 Z_{S^2} = \left(\frac r{r_0}\right)^{c/3} e^{-K_{tc}(\widetilde{\lambda},\overline{\widetilde{\lambda}})}
\label{MOeq:Conj}
\end{equation}
(where $r_0$ is a fixed scale),
as conjectured in \cite{MO2-sphere}.\footnote{The radial dependence
was suppressed in \cite{MO2-sphere}.}  This  quantity is
independent of scale and can be calculated
in the ultraviolet, for example on a gauged linear sigma model,
or directly in the infrared.

\section{Metrics on conformal manifolds}
\label{MOsec:metrics}

We now introduce two classes of $\mathcal{N}=(2,2)$ theories which give 
rise to conformal theories in the infrared.

\subsection{Nonlinear sigma models}
\label{MOsec:NLSM}

A nonlinear sigma model whose target is a Calabi--Yau manifold $X$ 
of complex dimension $n$ is a
2D quantum field theory with $\mathcal{N}=(2,2)$
supersymmetry which is expected to flow to a conformal theory 
of central charge $c=3n$ in
the infrared.  In fact, the $\beta$-function of such a theory vanishes
at one-loop, although there are in general higher loop 
corrections \cite{MOGrisaru:1986dk}.\footnote{In spite of these perturbative
corrections, one still expects 
a CFT in the infrared \cite{MONemeschansky:1986yx}.}
We choose the action of the $\mathcal{N}=(2,2)$ algebra on the
nonlinear sigma model in such a way that the chiral marginal operators
correspond to the harmonic $(n-1,1)$-forms on $X$, and the twisted
chiral marginal operators
correspond to the harmonic $(1,1)$-forms.
Thus, the chiral conformal manifold $\mathcal{M}_c$ corresponds to
the ``moduli space of $X$'' studied in algebraic geometry which specifies
the possible complex structures on $X$.  
The twisted chiral conformal manifold $\mathcal{M}_{tc}$, however,
 has no straightforward identification in mathematics.  Near the
``large radius limit'' boundary point it
is parameterized by the choice
of complexified K\"ahler form on $X$, which is a complex combination of
the K\"ahler form $\omega$ and the Kalb--Ramond two-form field $B$ (which
is only well-defined up to shifts by an integral two-form).
For this reason, $\mathcal{M}_{tc}$
is sometimes referred to as the ``complexified 
K\"ahler moduli space,'' with coordinate $t=i\omega+B$.

The Zamolodchikov metric on the chiral conformal manifold
$\mathcal{M}_c$ can be identified \cite{MOCandelas:1989qn} with
the Weil--Petersson metric which was described by Tian \cite{MOTian}
and Todorov \cite{MOTodorov-weil-petersson}.  For this description, we
consider the family of complex manifolds $\mathcal{X}\overset{\pi}{\to} \mathcal{M}_c$
corresponding to the variation of complex structure, and let $\Omega$
be a nonvanishing relative holomorphic $n$-form on $\pi^{-1}(U)$ over an open set 
$U\subset\mathcal{M}_c$.  Then the function
\begin{equation}
K_c:=- \log\left( i^{n^2} \! \! \int_X \Omega\wedge \overline{\Omega}\right)
\ ,
\end{equation}
which is a real-valued function on $U$, is a K\"ahler potential
for the Zamolodchikov metric restricted to $U$.
Any other choice of $\Omega$ takes the form $e^{-F}\Omega$
for  a nonvanishing holomorphic
function $e^{-F}$ on $U$.  If we make such a change, then
\begin{equation}
\begin{aligned}
- \log\left( i^{n^2} \! \! \int_X \Omega\wedge \overline{\Omega}\right)
&\mapsto
- \log\left( i^{n^2} \! \! \int_X e^{-F-\overline{F}} \Omega\wedge \overline{\Omega}\right)\\
&= - \log \left( i^{n^2} \! \! \int_X \Omega\wedge \overline{\Omega}\right) + F + \overline{F} \ ,
\end{aligned}
\end{equation}
as expected for a K\"ahler potential.  Due to some powerful non-renormalization theorems 
\cite{MOGrisaru:1979wc,MOSeiberg:1993vc}, this formula for the K\"ahler
potential on $\mathcal{M}_c$ is not subject to quantum corrections.

On the other hand, the twisted chiral conformal manifold
$\mathcal{M}_{tc}$ has a classical approximation in terms of the
K\"ahler cone $\mathcal{K}_X$ of $X$, complexified to 
$H^2(X,\mathbb{R}) + i \mathcal{K}_X$ by the inclusion of the
Kalb--Ramond field.  In the simplest case,\footnote{Other
cases can be handled by expressing $\mathcal{K}_X$ as a union of such
``integer basis'' cones up to automorphism; see \cite{MOcompact}.}
there are line bundles $\mathcal{L}_j$ whose first Chern classes
$c_1(\mathcal{L}_j)$ form a basis for $H^2(X,\mathbb{Z})$ and
also generate the K\"ahler cone:
\begin{equation}
\mathcal{K}_X = \mathbb{R}_{>0}\, c_1(\mathcal{L}_1) + \cdots +
\mathbb{R}_{>0}\, c_1(\mathcal{L}_s) .
\end{equation}
If $t_1$, \dots, $t_s$ are the corresponding complex coordinates on
$H^2(X,\mathbb{R})+i\mathcal{K}_X$, then $e^{2\pi it_1}$, \dots,
$e^{2\pi it_s}$ are local coordinates on the twisted chiral conformal
manifold.  With respect to these coordinates,
the K\"ahler potential for the
Zamolodchikov metric on $\mathcal{M}_{tc}$ has a classical 
expression
\begin{equation}
K_{tc} = 
-\log \left(\frac1{(2\pi)^n} \exp\left(\sum t_j\mathcal{F}_j\right) \wedge \overline{\exp\left(\sum {t}_j \mathcal{F}_j\right)} \right) + \cdots ,
\label{MOeq:classical-abstract}
\end{equation}
where
$\mathcal{F}_j$ is the curvature of a connection on the bundle $\mathcal{L}_j$,
expressed as a $2$-form (with indices suppressed), and the exponential
is computed as a power series in which differential
forms of even degree are multiplied
using the wedge product.

Using the fact that $L_j:=c_1(\mathcal{L}_j)$ is an integral cohomology class
represented by the differential form $\frac i{2\pi}\mathcal{F}_j$,
\eqref{MOeq:classical-abstract} can be rewritten in terms of integral
cohomology (evaluated on the fundamental homology class $[X]$) as
\begin{equation}
\begin{aligned}
e^{-K_{tc}} 
&= \left( \exp\left(\sum 2 \operatorname{Im}(t_j)L_j\right)\right)[X] + \cdots
\label{MOeq:classical-coho}
\end{aligned}
\end{equation}
where in this formula, the multiplication in the power series expansion is
represented by cup product.  Only the term in the exponential
of degree  $n$
 contributes to this
classical expression, which can be written as
\begin{equation}
e^{-K_{tc}} = 
\frac1{n!}
\left(\sum 2 \operatorname{Im}(t_j)L_j\right)^n[X] + \cdots
\label{MOeq:singleterm}
\end{equation}
and depends on the 
intersection pairings among the integral divisors
$L_j$,
which
are specified by the 
cohomology
ring of $X$.

In either form,
this classical
expression is subject to both perturbative and non-perturbative corrections,
to be discussed in the next section.

\subsection{Gauged linear sigma models}
\label{MOsec:GLSM}

Another approach to  conformal field theories on  Calabi--Yau
manifolds is to start with a Lagrangian theory in the UV known as
a {\em gauged linear sigma model}\/ \cite{MOWitten:1993yc}.

A gauged linear sigma model (GLSM)
is formulated in $\mathcal{N}=(2,2)$
superspace, and involves a compact gauge group $G$ 
as well as $N$ chiral matter multiplets
transforming in a representation $\Psi:G\to U(N)$.
We denote the
corresponding representation 
of the Lie algebra $\mathfrak{g}$ of $G$ by
$\psi:\mathfrak{g}\to\mathfrak{u}(N)$,
so that\footnote{We follow the usual
physics convention of putting an $i$ in the exponential map so that
the Lie algebra consists of Hermitian operators.  We also put a factor
of $2\pi$ to clarify the integral structure.}
\begin{equation}
\Psi(e^{2\pi iY})=e^{2\pi i\psi(Y)}
\end{equation}
for $Y\in\mathfrak{g}$.
To streamline our later analysis,
we fix a Cartan subgroup $H\subset G$ (i.e., a maximal connected
abelian subgroup) with corresponding Cartan subalgebra
$\mathfrak{h}\subset \mathfrak{g}$,
and choose coordinates $\phi_J$ on the
complex vector space $\mathbb{C}^N$ such that 
each $\phi_J$ is a simultaneous eigenvector for $\Psi|_H$.
The eigenvalues of $\Psi_H$ can be specified by means of the {\em weight lattice}\/
$\Lambda_{\text{wt}}\subset \mathfrak{h}^*$ of $G$, which gives the eigenvalues for the
corresponding representation of $\mathfrak{h}$.  
That is, for each $\phi_J$ there is a weight vector 
$w_J\in \Lambda_{\text{wt}}\subset
\mathfrak{h}^*$ such that for $h=e^{2\pi i Y}\in H$,
\begin{equation}
\Psi(h)(\phi_J) = e^{2\pi i w_J\cdot Y}\phi_J ,
\end{equation}
using a dot to denote the pairing between $\mathfrak{h}^*$ and
$\mathfrak{h}$.

One of the interaction terms in the Lagrangian is specified
by means of a $G$-invariant ``superpotential'' polynomial
$W(\phi_1,\dots,\phi_N)$.
We will also construct a term in the Lagrangian from Lie algebra characters
$\xi:\mathfrak{g}\to \mathfrak{u}(1)$ which arise from one-dimensional
representations
$\Xi:G\to U(1)$. It is convenient to choose a basis
$\xi_1$, \dots, $\xi_k$ for the lattice of such characters.  All of these
characters are trivial on the commutator $[\mathfrak{g},\mathfrak{g}]$ and 
so factor through the projection to 
the abelianization $\mathfrak{a}=\mathfrak{g}/[\mathfrak{g},\mathfrak{g}]$.

To construct the GLSM, we begin with $N$ chiral
superfields $\Phi_J$ (i.e., satisfying $\overline{D}_+\Phi_J =
\overline{D}_-\Phi_J=0$) interacting via the holomorphic superpotential
$W(\Phi_1,\ldots,\Phi_N)$.  The model is invariant under the action of
$G$ (via $\Psi$) on $\mathbb{C}^N$ and we gauge this action, preserving 
$\mathcal{N}=(2,2)$
supersymmetry, by introducing a ${\mathfrak g}$-valued vector multiplet
$V$ with invariant field strength
$\Sigma=\frac1{\sqrt{2}}\overline{D}_+D_-V$.  This last field is
{\it twisted chiral}, which means that $\overline{D}_+\Sigma
=D_-\Sigma=0$.  
We include a topological theta angle $\vartheta$ and a Fayet--Iliopoulos
D-term with coefficient $\zeta$,
each taking values in\footnote{Globally, as discussed in 
\cite{MOHori:2011pd,MOHori:2013gga,MOKnapp:2016rec},
we should identify $e^{-2\pi\zeta+i\vartheta}$ with an element of
$\operatorname{Hom}(\pi_1(G),\mathbb{C}^*)^{G_0}$,
where $G_0$ is the connected component of $G$.} 
$\mathfrak{a}^*=
\operatorname{Ann}([\mathfrak{g},\mathfrak{g}])\subset\mathfrak{g}^*$.
These
terms are naturally written in terms of the complex combination\footnote{Both
$\tau$ and $z$
are local coordinates on the twisted chiral conformal manifold.  To avoid
cluttering our formulas, we suppress the ``tilde'' on these variables
which should be present
for consistency with the notation of
Section~\ref{MOsec:KpZ}.}
$\tau=i\zeta+\frac
1{2\pi}\vartheta$ or its exponential $z=e^{2\pi i\tau}=e^{-2\pi \zeta + i\vartheta}$.  We introduce a pairing between $\mathfrak{a}^*$ and $\mathfrak{g}$ 
defined by
\begin{equation}
\langle \tau, \Sigma \rangle
:= \sum \tau_a \operatorname{tr}_{\xi_a}(\Sigma) ,
\end{equation}
which is independent of the choice of basis.
The resulting Lagrangian density is 
\begin{eqnarray}
{\cal L}&=&\int d^4\theta\left(\|e^{\psi({V})}{\Phi}\|^2
-\frac1{4e^2}\|{\Sigma}\|^2\right)\nonumber\\
&&+ \left(\int d\theta^+d\theta^- W(\Phi_1,\dots,\Phi_N) +
 \mbox{ c.c. } \right)\label{MOGLSM}\\
&&+ \left( {\frac i{\sqrt 2}}\,
\int d\theta^+d\bar\theta^- \langle \tau,\Sigma\rangle
+ \mbox{ c.c. } \right) ,\nonumber
\end{eqnarray}
where for simplicity of notation, we have set all gauge couplings of irreducible
factors of $G$ to a single value $e$.
The marginal 
couplings of these theories
are the coefficients\footnote{More precisely, the coefficients
account for the marginal chiral couplings with some redundancy; see
\cite{MOmondiv} or \cite{MOtowards-duality}  for an account of this.}
 of the superpotential
(for $\mathcal{M}_c$) and the choice of D-term coefficient and $\theta$-angle
(for $\mathcal{M}_{tc}$).

We will assume that these theories admit both a vector-like symmetry 
$U(1)_V$ and an axial-like symmetry $U(1)_A$.  
In flat space, a given action of $U(1)_V\times U(1)_A$ may be modified
by a global symmetry but on $S^2$, 
changing the charges
of the fields (other than by adding gauge charges)
produces a distinct theory  \cite{MOBenini:2012ui, MODoroud:2012xw}.  
For this reason, we shall regard
the specification of these charges as part of the data of the theory.
We design our choice with the expectation that, should the theory
flow to a superconformal theory in the IR, the specified $U(1)_V\times U(1)_A$
will become the $R$-symmetry which is part of the superconformal algebra.
In general we allow these $R$-charges to be rational numbers, although
for many purposes it is best if they are integers up to gauge transformation.
We will study this theory on $S^2$ preserving the $U(1)_V$ symmetry,
in order to analyze the metric on the twisted chiral conformal manifold.

Having specified an $R$-charge $q\in \mathbb{Q}$ for a given superfield, the general form
of the vector-like symmetry is
\begin{equation}e^{i\alpha F_V}:\mathcal{F}(x^\mu,\theta^\pm,\bar\theta^\pm)
\mapsto e^{i\alpha q} \mathcal{F}(x^\mu,e^{-i\alpha}\theta^\pm,e^{i\alpha}
\bar\theta^\pm)\end{equation}
while the general form of the axial-like symmetry is
\begin{equation}e^{i\beta F_A}:\mathcal{F}(x^\mu,\theta^\pm,\bar\theta^\pm)
\mapsto e^{i\beta q} \mathcal{F}(x^\mu,e^{\mp i\beta}\theta^\pm,e^{\pm i\beta}
\bar\theta^\pm)\ . \end{equation}
Note that the superpotential, if nonzero, must have $R$-charge $2$.

In flat space, such a symmetry can have an anomaly in the presence of
gauge fields.  
A quick computation \cite{MOWitten:1993yc,MOsumming} shows that
the anomaly is given by a function on the Lie
algebra proportional to $V\mapsto\operatorname{tr}(\psi(V))$;
we require that this vanish identically so that the symmetries are not
anomalous.
Since the action of the continuous part
of $G$ on the monomial $\Phi_1\cdots \Phi_N$ is via
$\exp(\operatorname{tr}(\psi(V)))$, this is the same as requiring
that $\Phi_1\cdots \Phi_N$
be invariant under the continuous part of the gauge group $G$.
In addition, in order to ensure integral $R$-charges up to gauge
transformation, we require that $\Phi_1\cdots \Phi_N$
 be invariant under the entire group $G$
\cite{MOAspinwall:2015zia}.
In other words, we must require that 
\begin{equation}
\Psi(G)\subset SU(N) ,
\label{MOeq:anomalyfree}
\end{equation} 
and we will impose that requirement henceforth.

Finally, using the $R$-symmetry and
calculating as in \cite{MOSilverstein:1994ih}, one finds that
the central charge $c$ of the fixed-point CFT is determined by
\begin{equation}
\frac c3 = d-\sum_{J=1}^Nq_J\ ,
\end{equation}
where the sum extends over all of the chiral superfields,
and where $d$ is the difference between the
number of chiral fields and the number of
gauge fields.

\subsection{Phases of an abelian GLSM}
\label{MOsec:abelian-phases}
\label{MOsec:example}

As explained in detail in \cite{MOWitten:1993yc}, a GLSM with an abelian
gauge group and an anomaly-free $R$-symmetry (i.e., $\sum w_J=0$)
can be described very explicitly at low energy and in many
cases coincides with a nonlinear sigma model with target a Calabi--Yau
manifold.  In such cases, the correspondence only holds when the FI-parameters
are in a certain range of values, and the typical
GLSM has other phases with different
low-energy descriptions in addition to the geometric phase(s).

The low-energy analysis begins by 
mapping out the space of
classical vacua of the theory.  
The algebraic
equations of motion for the auxiliary fields $D_a$ in the vector
multiplets and $F_J$ in the chiral multiplets can be solved:
\begin{eqnarray}
D_a &=& -e^2\left(\sum_{J=1}^N (w_J)_a |\phi_J|^2 - \zeta_a\right)\label{MOD}\\
F_J &=& -{\frac{\partial W}{\partial\phi_J}}\ .\label{MOF}
\end{eqnarray}
The potential energy for the bosonic zero modes is then
\begin{equation}
U = {\frac1{ 2e^2}}\sum_{a=1}^{h}D_a^2 + \sum_{J=1}^N |F_J|^2 +
\sum_{a,b}\bar\sigma_a\sigma_b\sum_{J=1}^N (w_J)_a(w_J)_b|\phi_J|^2\ ,
\end{equation}
where $\phi_J,\,\sigma_a$ are the lowest components of
$\Phi_J,\,\Sigma_a$ respectively.  The space of classical vacua is the
quotient by $G$ of the set of zeros of $U$.

Suppressing
any solutions with $\sigma\ne0$ (which are absent for generic
values of the parameters), the space of solutions
 is 
\begin{equation}
(D^{-1}(0) \cap \operatorname{Crit}(W))/G \subset
D^{-1}(0)/G .
\end{equation}
Since $G$ is abelian,
this quotient has a description as
a toric variety of dimension $N-h$, as we now review.
The
group $G$ has a natural complexification $G_{\mathbb{C}}$ in which
each $U(1)$ factor is promoted to the complex group $\mathbb{C}^*$;
 the action of $G$ on $V$ extends to an action of $G_{\mathbb{C}}$.
For any choice of FI-parameters, the space $D^{-1}(0)/G$ has a description
as a GIT quotient 
\begin{equation}
\mathbb{C}^N\mathbin{/\!/}G_{\mathbb{C}}
= (\mathbb{C}^N - Z_{\zeta})/G_{\mathbb{C}},
\end{equation}
where $Z_\zeta$ is the
union of all $G_{\mathbb{C}}$-orbits which do not meet meet
$D^{-1}(0)$.  (The dependence on the FI-parameters $\zeta$ comes from
their inclusion in the D-term equation \eqref{MOD}).

The nature of the quotient space changes as $\zeta$ varies, and can
be systematically described by a construction known as the
``secondary fan''  \cite{MOMR1020882,MOMR1073208,MOBFS,MOcatp}. 
Let $\mathcal{J}\subset \{1, \dots, N\}$ be a collection of $h$ indices
such that the weight vectors $\{ {w}_J, J\in \mathcal{J}\}$ are linearly
independent; we wish to know if $D^{-1}(0)$ contains entries
in which 
\begin{equation}
\phi_J=0 \text{ for all } J\not\in \mathcal{J} .
\label{MOeq:phi-vanish}
\end{equation}
(If so, then the corresponding
orbit $\mathcal{O}_{\mathcal{J}}$ lies in the quotient 
space.\footnote{A bit more concretely, for each $\mathcal{G}_{\mathcal{J}}$
containing $\zeta$, the complementary set of indicies
$\{1,\dots,N\}-\mathcal{J}$ labels the coordinates for one of the
toric cooridnate charts of $D^{-1}(0)/G$.})
To answer this, note that if we impose \eqref{MOeq:phi-vanish} then
the D-term equations become
\begin{equation}
\zeta = \sum_{J\in \mathcal{J}} |\phi_J|^2 {w}_J .
\label{MOeq:cone}
\end{equation}
In other words, since all coefficients on the right side of \eqref{MOeq:cone}
are nonnegative, $\zeta$ must lie in the cone $\mathcal{C}_{\mathcal{J}}$
in FI-parameter-space $\mathfrak{a}^*$ which
is generated by the weight vectors $\{{w}_J, J\in \mathcal{J}\}$.

Thus, for a given $\zeta\in\mathfrak{a}^*$,
to determine which orbits $\mathcal{O}_{\mathcal{J}}$ lie in
$D^{-1}(0)/G$ we simply determine which cones $\mathcal{C}_{\mathcal{J}}$ contain
$\zeta$.  This decomposition into cones describes the {\em secondary
fan}, and the regions it defines in $\mathfrak{a}^*$ are called
the {\em phases}\/ of the GLSM.

\begin{figure}[t]
\centering
\includegraphics[height=2.8in]{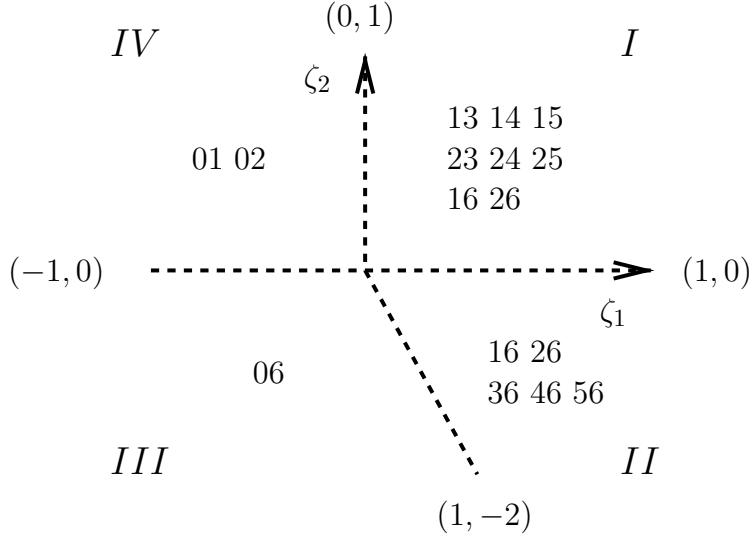}
\caption{Secondary fan data for the example.}\label{MOfig:secondary}
\end{figure}

To illustrate this construction, we work it out in a particular 
example\footnote{The example we use is an abelian GLSM, which does not
exhibit the full power of the localization method to compute Gromov--Witten
invariants, since those invariants can also be computed by mirror symmetry
for abelian GLSMs with a geometric phase.  However, we avoid some of
the complications of nonabelian GLSMs by working with this particular example.}
which we will follow throughout the paper.  
We use an example 
which has been studied extensively in the 
literature \cite{MO2param1,MOHosono:1993qy,MOsmall},\cite{MOsumming},\cite{MOlooking,MOhhp,MOHori:2013ika}.

Consider an anti-canonical hypersurface in the toric variety obtained
from the weighted projective fourfold $\mathbb{P}^{(1,1,2,2,2)}$
by blowing up its singular locus.
This can be described by a GLSM as follows.
Let $(\gamma_1,\gamma_2)\in G = U(1)\times U(1)$ act on the vector space $\mathbb{C}^7$ via
\begin{equation}
  (\phi_0,\phi_1,\dots,\phi_6) \mapsto 
(\gamma_1^{-4}\,\phi_0,\gamma_2\,\phi_1,\gamma_2\,\phi_2,\gamma_1\,\phi_3,
  \gamma_1\,\phi_4,\gamma_1\,\phi_5,\gamma_1\gamma_2^{-2}\,\phi_6)\ .
\end{equation}
We specify $R$-charges of these fields in terms of two arbitrary
rational parameters ${q}_1$ and ${q}_2$ to be determined later, as
\begin{equation} \label{MOeq:RDeg8}
\begin{tabular}{|l|c|c|c|c|} \hline
field & $\phi_0$ & $\phi_1$, $\phi_2$ & $\phi_3$, $\phi_4$, $\phi_5$ & $\phi_6$
\\ \hline
$R$-charge & $2-4q_1$ & $q_2$ & $q_1$ & $q_1-2q_2$ 
\\ \hline
\end{tabular}
\end{equation}
For the superpotential, which is a $G$-invariant polynomial of $R$-charge two, we choose
\begin{equation}
 W(\phi_0,\phi_1,\ldots,\phi_6) = \phi_0 \, F_{(4,0)}(\phi_1,\ldots,\phi_6),
\label{MOeq:superpot}
\end{equation}
where $F_{(4,0)}$ is a generic
homogeneous polynomial of bi-degree $(4,0)$ with respect
to the gauge group $G=U(1)\times U(1)$.

The D-term
equations are
\begin{align}
\label{MOeq:moment1}
 \zeta_1 &= -4|\phi_0|^2 + |\phi_3|^2+|\phi_4|^2+|\phi_5|^2+|\phi_6|^2 \ ,\\
 \zeta_2 &= |\phi_1|^2+|\phi_2|^2-2|\phi_6|^2 \ .
\label{MOeq:moment2}
\end{align}
We then find the secondary fan data which is illustrated in
Figure~\ref{MOfig:secondary}:  for each pair $\{J, J'\}$ we have
indicated the region(s) which are included in the cone generated by
${w}_J$ and ${w}_{J'}$.  This leads to four phases, labeled
by Roman numerals in the Figure.

Note that the cones $\mathcal{C}_{\mathcal{J}}$ are not necessarily phase regions in and of
themselves; in the example, $\mathcal{C}_{16}$ is the union of phases I and II.

The geometry of the various quotients is best described by determining the
sets $Z_{\zeta}$ which are excluded from the quotient.  If we label
those sets according to phase region, then by examining
which variables are allowed to vanish together, we find that 
\begin{align}
Z_{I} &= \{\phi_1=\phi_2=0\} \cup \{ \phi_3=\phi_4=\phi_5=\phi_6=0\} \\
Z_{II} &= \{ \phi_1=\phi_2=\phi_3=\phi_4=\phi_5=0\} \cup \{\phi_6=0\}\\
Z_{III} &= \{\phi_0=0\} \cup \{\phi_6=0\} \\
Z_{IV} &= \{ \phi_0=0\} \cup \{\phi_1=\phi_2=0\}
\end{align}
Each phase has a geometric description \cite{MOsumming}:  in phase I, we get a line bundle over the blowup of $\mathbb{P}^{(1,1,2,2,2)}$ along its singular locus, in phase II, we get a line bundle over $\mathbb{P}^{(1,1,2,2,2)}$ itself,
in phase III we get $\mathbb{C}^5/\mathbb{Z}_8$, and in phase IV we get
\begin{equation}
 \left(\mathbb{C}^3 \times \mathcal{O}_{\mathbb{P}^1}(2)\right)/\mathbb{Z}_4.
\end{equation}
The phase of relevance for comparison to the nonlinear sigma model is the
geometric phase, phase I.  

We still must impose the F-term equations, and in doing so, we can
be more specific concerning our ``generic'' assumptions about the superpotential \eqref{MOeq:superpot}.
We assume that
$F_{(4,0)}(\phi_1,\phi_2,\phi_3,\phi_4,\phi_5,1)$ is a
transverse homogeneous polynomial in $5$ variables (which means that the origin is
the only common zero of the partial derivatives), and that
$F_{(4,0)}(\phi_1,\phi_2,\phi_3,\phi_4,\phi_5,0)$, which is independent
of $\phi_1$ and $\phi_2$, is a transverse homogeneous polynomial of
$3$ variables.  The F-term equations are
\begin{equation}
\frac{\partial W}{\partial \phi_J} = 0, \quad J=0,\dots,6 .
\end{equation}

To solve the F-term equations, we note that $\partial W/\partial \phi_0=F$,
and that $\phi_0$ divides $\partial W/\partial \phi_J$ for $J\ne0$.
Thus, one solution is 
\begin{equation}
\phi_0=F=0.
\label{MOeq:sol1}
\end{equation}
  If $\phi_0\ne0$ but $\phi_6=0$,
then $(\partial F/\partial \phi_j)|_{\phi_6=0}=0$ for $J=3,4,5$, which
implies (by transversality) that 
\begin{equation}
\phi_3=\phi_4=\phi_5=\phi_6=0.
\label{MOeq:sol2}
\end{equation}
  Moreover, there is no monomial
appearing in $F$ which involves only the variables $\phi_1$ and
$\phi_2$,  so $F$ vanishes on
the locus \eqref{MOeq:sol2} and we see that \eqref{MOeq:sol2} provides
a second solution to the F-term equations.  Finally, if $\phi_0\ne0$ and
$\phi_6\ne0$, then the transversality of 
$F(\phi_1,\phi_2,\phi_3,\phi_4,\phi_5,1)$ implies that
\begin{equation}
\phi_1=\phi_2=\phi_3=\phi_4=\phi_5=0
\label{MOeq:sol3}
\end{equation}
is also a solution (because there is no power of $\phi_6$ is a monomial).

In phase I, we see that that only possible solution is $\{\phi_0=F=0\}$.
This defines a hypersurface $F=0$ inside the zero-section $\{\phi_0=0\}$
of the line bundle.  In other words, we get a hypersurface in
the blowup of $\mathbb{P}^{(1,1,2,2,2)}$ along its singular locus.
The requirement \eqref{MOeq:anomalyfree} (guaranteeing anomaly-free $R$-symmetry
and integral $R$-charges up to gauge transformation) is precisely
the condition for the hypersurface to be Calabi--Yau.  This is a
general phenomenon for toric hypersurfaces \cite{MOBatyrev1993}.

The other solutions to the F-term equations are relevant in other
phases:  in phase II, we again get \eqref{MOeq:sol1}; in phase III, we
get \eqref{MOeq:sol3}; and in phase IV, we get \eqref{MOeq:sol2}.

\section{Quantum corrections to the K\"ahler potential}
\label{MOsec:perturb}

In this section, we discuss quantum effects in the $\mathcal{N}=(2,2)$
supersymmetric two-dimensional nonlinear $\sigma$-model on a Calabi--Yau
manifold $X$ of arbitrary dimension.  
As noted above, the K\"ahler potential $K_c$ on the chiral conformal manifold
is not subject to corrections.  However, the K\"ahler potential $K_{tc}$ on 
the twisted chiral conformal manifold is subject to perturbative corrections 
which have been
determined in detail in \cite{MOgamma-class}, as well as non-perturbative
instanton corrections described in terms 
of Gromov--Witten invariants.  
We describe the perturbative corrections both in terms of expressions in
the Riemannian curvature (integrated over $X$) and in terms of cohomology classes
(evaluated on the fundamental homology class of $X$).

\subsection{Nonlinear $\sigma$-model action and the effective action}

Under the renormalization group an $\mathcal{N}=(2,2)$ supersymmetric, two-dimensional, nonlinear $\sigma$-model with K\"ahler target space $X$ (of complex dimension $n$), flows in the infrared to a conformal fixed point characterized by vanishing $\beta$-functions. In this section, the $\beta$-function of the target space K\"ahler form is of particular interest, which vanishes at tree level but is nonzero at one-loop:
\begin{equation} \label{MOeq:betaij}
   \frac{1}{\alpha'} \beta_{i\bar\jmath}\,=\, R_{i\bar\jmath} + \Delta\omega_{i\bar\jmath}(\alpha') \,=\,R_{i\bar\jmath}+ {\alpha'}^3 \frac{\zeta(3)}{48} T_{i\bar\jmath} + O({\alpha'}^5) \ .
\end{equation}   
Here $\alpha'$ is the coupling constant in the nonlinear $\sigma$-model. At leading one-loop order, the Ricci tensor $R_{i\bar\jmath}$ appears; $\Delta\omega_{i\bar\jmath}$ then comprises all higher loop corrections, which are exact in cohomology, i.e., $\Delta\omega = d\rho$ with some global one form $\rho$ on $X$ \cite{MOHowe:1986ys,MONemeschansky:1986yx}.
 The tensor $T_{i\bar\jmath}$ is the first non-vanishing subleading correction at four loops \cite{MOGross:1986iv}, which has been explicitly calculated in \cite{MOGrisaru:1986px}. (The five-loop correction at order $O({\alpha'}^4)$ has been shown to vanish \cite{MOGrisaru:1986wj}.)
Thus, at leading order the vanishing $\beta$-function $\beta_{i\bar\jmath}=0$ requires a Ricci-flat K\"ahler metric and hence a Calabi--Yau target space. However, this Ricci-flat Calabi--Yau target space metric gets further corrected at higher loops. 

To analyze these corrections, it is useful to adopt an effective action point of view for the target space geometry. Namely, we interpret the condition for the vanishing $\beta$-function as the Euler--Lagrange equation for the metric $g_{i\bar\jmath}$ arising from an action functional \cite{MOGross:1986iv,MOZanon:1986gg}. The relevant effective action $\mathcal{S}_{\rm eff}[g]$ takes the form
\begin{equation} \label{MOeq:EffAction}
  \mathcal{S}_{\rm eff}[g]\,=\, \int \sqrt{g} \left[ R(g) + \Delta S(\alpha',g) \right] \ ,
\end{equation}  
with the corrections $\Delta S(\alpha',R)$. The leading correction arises at fourth loop order ${\alpha'}^3$ and enjoys the expansion
\begin{equation}
  \Delta S(\alpha',g) \,=\, {\alpha'}^3 S^{(4)}(g) + {\alpha'}^5 S^{(6)}(g) + \ldots \ .
\end{equation}
Here the $n$-th loop correction $S^{(n)}(g)$ is a scalar functional of the metric tensor and the Riemann tensor. A proposal for the structure of these terms 
was  put forward in \cite{MOFreeman:1986zh}. 

Since the effective action $\mathcal{S}_{\rm eff}[g]$ gets corrected beyond the leading contribution, we expect the classical metric on $\mathcal{M}_{tc}$ to receive further corrections from higher loop orders. 
Using mirror symmetry, 
the tree level term and four-loop correction 
in the case of Calabi--Yau threefolds 
were determined to be \cite{MOCDGP}
\begin{equation} \label{MOeq:K4loop}
\begin{aligned}
  e^{-K_{tc}}\,&=\, \frac{1}{3!}  \left(  \sum_{j=1}^s \! 2 \operatorname{Im} t_j\, L_j \! \right)^3 \!\! [X]
  + {\alpha'}^3
\left(-\frac1{4\pi^3}\zeta(3)c_3(X) \cup \left( \sum_{j=1}^s \! 4 \pi \operatorname{Im} t_j\, L_j \! \right)^{0}  \right)[X] + O(e^{2\pi i t}) \ ,
\end{aligned}
\end{equation}
expressed
in terms of the  Chern class  $c_3(X)$ and a special value of the Riemann $\zeta$-function. The appearance of the $\zeta$-value $\zeta(3)$ (of transcendental weight three) indicates its origin as a four-loop counterterm of the $\mathcal{N}=(2,2)$ supersymmetric nonlinear $\sigma$-model \cite{MOGrisaru:1986px}.

In general, further corrections in ${\alpha'}$ appear for Calabi--Yau target spaces of higher dimension $n>3$. They take the following form in which $\alpha'$ is indicated
explicitly (although it is set equal to $1$ elsewhere in this review):
\begin{equation} \label{MOeq:ChiClasses}
  e^{-K_{tc}}\,=\, \left(  \exp\!\left(\!{ \sum_{j=1}^s 2 \operatorname{Im} t_j\,L_j} \!\right)+   \frac1{(2\pi)^n}\sum_{k=0}^n {\alpha'}^k \chi_k \right)[X] +  O(e^{2\pi i t}) \ ,
\end{equation}
where the characteristic class $\chi_k$ arises from the perturbative loop corrections at loop order $k+1$. 
(Thanks to \cite{MOGrisaru:1986wj} we expect $\chi_4$ to vanish.)
Due to the appearance of higher curvature tensors in the corrections $\Delta\omega_{i\bar\jmath}$ of the $\beta$-function~\eqref{MOeq:betaij}, 
we can expect that integrating such curvature tensors can be expressed in terms of the Chern classes of the tangent bundle of the target space~$X$. Furthermore, the loop corrections appearing in $\Delta\omega_{i\bar\jmath}$ at a given loop order $k+1$, i.e., at order ${\alpha'}^{k}$, give rise to corrections with transcendentality degree $k$, which  is a general property of loop corrections of supersymmetric two-dimensional $\sigma$-models \cite{MOBroadhurst:1996ur}. As a result, the cohomology classes $\chi_k$ are homogeneous elements of transcendental degree $k$ in the graded polynomial ring over all products of multiple $\zeta$-values up to transcendental weight $k$
\begin{equation} \label{MOeq:PClass}
   \chi_k \in H^{2k}(X,\mathbb{Q})[\zeta(m)_{2\le m\le k}, \zeta(m_1,m_2)_{2\le m_1+m_2 \le k},\ldots,\zeta(1,\ldots,1)]_k \ .
\end{equation}
The transcendental weight of a multiple $\zeta$-value $\zeta(m_1,\ldots,m_a)$ is given by the sum $m_1+\ldots+m_a$, and the multiple zeta functions $\zeta(m_1,\ldots,m_a)$ generalize the Riemann zeta function according to \cite{MOMR1341859}
\begin{equation}
  \zeta(m_1,\ldots,m_a) \,=\, \sum_{n_1>n_2>\ldots>n_a} \frac1{n_1^{m_1} \cdots n_a^{m_a}} \ .  
\end{equation}
Note that there are many non-trivial relations over $\mathbb{Q}$ among such multiple $\zeta$-values; see for instance~\cite{MOMR2578167}.

\subsection{Perturbative corrections to the K\"ahler potential} \label{MOsec:proposal}

Perturbative corrections to $e^{-K_{tc}}$ were found in \cite{MOgamma-class},
assuming that all corrections take the universal form \eqref{MOeq:ChiClasses},
by using mirror symmetry and period computations to determine the 
values of $\chi_k$.
The answers can be
expressed in terms of a characteristic class known as the ``gamma class'' 
\cite{MOarXiv:math.AG/9803119,MOarXiv:0712.2204,MOMR2553377,MOarXiv:1101.4512,MOMR2483750},\cite{MOHori:2013ika},
but we will take a more direct approach
and present the corrections explicitly
both in terms of Riemannian curvature and in terms
of Chern classes.

The corrections to the K\"ahler potential involve the Riemann curvature tensor, which we denote
by $\mathcal{R}$ (suppressing indices) and regard as a differential-form-valued
 endomorphism of the tangent bundle of $M$.  If we take the trace over
tangent bundle indices, we obtain a $2$-form $\operatorname{tr}(\mathcal{R})$
which is just the familiar Ricci tensor $R_{ij}$ with indices suppressed.  
We will also consider traces of higher powers (i.e., composing the endomorphism
with itself a number of times):  $\operatorname{tr}(\mathcal{R}^\ell)$ defines a $2\ell$-form.

We can now state the perturbative corrections to the classical metric 
\eqref{MOeq:classical-abstract} which were derived in \cite{MOgamma-class},
in the case of $c_1(X)=0$:
\begin{equation}
\begin{aligned}
K_{tc} &= 
-\log \frac1{(2\pi)^n}\int_X \exp\left(2\operatorname{Re}\left(\sum t_j\mathcal{F}_j\right) 
 -2 \sum_{k=1}^\infty
\frac{\zeta(2k+1)}{2k+1} \operatorname{tr}\left( (i\mathcal{R}/2\pi )^{2k+1}\right) \right) \\
& \qquad + \mathcal{O}(e^{2\pi it}) \ .
\end{aligned}
\label{MOeq:perturbative-with-F}
\end{equation}

To write this in terms of integer cohomology classes, we need to use
Newton's identities which express $\sum_{j=1}^n x_j^\ell$
in terms of the elementary symmetric functions $\sigma_1$, $\sigma_2$, \dots,
$\sigma_n$ of $\{x_1,x_2,\dots,x_n\}$.  If we write
\begin{equation}
\sum_{j=1}^n x_j^\ell = P_\ell(\sigma_1,\sigma_2,\dots,\sigma_\ell) \ ,
\label{MOeq:Newton}
\end{equation}
then by Chern--Weil theory, 
\begin{equation}
\int_X \operatorname{tr}\left(( i\mathcal{R}/2\pi)^\ell\right) = P_\ell(c_1,c_2,\dots,c_\ell) \ .
\end{equation}
Thus, we can express the perturbative corrections to \eqref{MOeq:classical-coho}
in the form
\begin{equation}
\begin{aligned}
e^{-K_{tc}} 
&= \exp\left(\sum 2 \operatorname{Im}(t_j)L_j
 -\frac2{(2\pi)^n} \sum_{k=1}^\infty \frac{\zeta(2k+1)}{2k+1}
P_{2k+1}(c_1,c_2,\dots,c_{2k+1}) \right)
[X] \\
& \qquad + \mathcal{O}(e^{2\pi it}) \ ,
\label{MOeq:perturbative-coho}
\end{aligned}
\end{equation}
where $c_1=0$, $c_2$, $c_3$, \dots \  are the Chern classes of $X$.

For Calabi--Yau threefolds, the perturbative correction \eqref{MOeq:K4loop}
found in \cite{MOCDGP} gives the complete answer.  However,
in order to evaluate these expressions for Calabi--Yau manifolds
of higher dimension, it is convenient to have
the first several (odd) Newton's identities at our disposal, which
we give with $\sigma_1$ set to $0$:
\begin{equation}
\begin{aligned}
P_1|_{\sigma_1=0} &= 0\\
P_3|_{\sigma_1=0} &= 3\sigma_3 \\
P_5|_{\sigma_1=0} &= 
-5\sigma_2\sigma_3+5\sigma_5\\
P_7|_{\sigma_1=0} &= 
7\,{\sigma_{{2}}}^{2}\sigma_{{3}}
-7\,\sigma_{{2}}\sigma_{{5}}
-7\,\sigma_{{3}}\sigma_{{4}}
+7\,\sigma_{{7}}
\end{aligned}
\label{MOeq:newton-list-simplified}
\end{equation}

Setting $c_1=0$ and expanding the exponential in 
\eqref{MOeq:perturbative-coho}, we find
the first few perturbative corrections 
\begin{equation} \label{MOeq:first-few}
\begin{aligned}
   \chi_3\,&=\,-2 \zeta(3) c_3 \\
 \chi_4\,&=\,0 \\
\chi_5\,&=\,2 \zeta(5) \left( c_2 c_3 -  c_5 \right) \\
   \chi_6\,&=\, 2 \zeta(3)^2 c_3^2 \\
\chi_7\,&=\, -2 \zeta(7) \left(c_2^2 c_3 -c_3c_4-c_2c_5+c_7\right) \\
\end{aligned}
\end{equation}  
in terms of the Chern classes $c_k$ of the Calabi--Yau $n$-fold $X$.
The contributions $\chi_3$ and $\chi_4$ are exactly what appear in
\cite{MOCDGP,MOarXiv:1302.3760}.

\subsection{Nonperturbative corrections and Gromov--Witten invariants}
\label{MOsec:np}

The nonperturbative corrections to a two-dimensional nonlinear sigma model
are due to instantons, i.e., action-minimizing maps
from a Euclidean spacetime to the target manifold, and the relevant
corrections to the metric on $\mathcal{M}_{tc}$ are given by
instantons of genus zero.

In order to describe the instanton corrections to the Zamolodchikov
metric, we must first describe instanton corrections to certain other
quantities in the theory.
The twisted chiral operators in the theory have a natural ring
structure determined by the two-point and three-point genus zero
correlation functions
in terms of a ``Frobenius algebra'' structure
\cite{MOMR1397274}.  This determination  goes as follows:
given a ring $R$ with a nondegenerate bilinear form
\begin{equation}
( \text{{--}}, \text{{--}} ): R\times R \to \mathbb{C},
\label{MOeq:bilinear}
\end{equation}
there is a natural trilinear map
\begin{equation}
\langle ABC \rangle := (A\star B, C),
\label{MOeq:trilinear}
\end{equation}
where $\star$ denotes the product in the ring.  (When evaluated on a basis
of $R$, this gives the structure constants for the ring.)
Conversely,   whenever we are given
a bilinear form \eqref{MOeq:bilinear} and a trilinear form
\eqref{MOeq:trilinear}, we get a product on $R$.

These two-point and three-point correlation functions
can be computed equally well in the closely related ``topological
sigma model'' \cite{MOtsm},
in which the spins of the fields are modified
and a
suitable projection is performed.\footnote{We
are considering here the ``A-model'' of \cite{MOWitten:1991zz}.} 
The twisted chiral operators in
the topological theory can be identified with  harmonic forms
(or their cohomology classes)
on the target manifold $X$, and the ring structure in the classical theory
 is simply the cup product pairing in cohomology.
However, this ring structure is deformed by instanton 
contributions \cite{MOtsm,MOtopgrav,MOVafa:1991uz} in the quantum theory,
giving rise to the
``quantum cohomology ring'' of $X$ (which is known to be
associative \cite{MOMR1144529,MODijkgraaf:1990dj,MODubrovin:1992dz}).

To describe the instanton contributions,
we represent cohomology classes $A$, $B$, and $C$ by algebraic cycles
on $X$; the correlation function is calculated by integrating
over the space of maps $\pi$ from the genus
one spacetime
with three points $p$, $q$, $r$ specified such that 
$\pi(p)$ lies in $A$, $\pi(q)$ lies in $B$,
and $\pi(r)$  lies in $C$; a standard localization procedure
reduces the computation to determining the space of volume-minimizing maps.  
The classical contribution to the 
correlation function comes from constant maps, and simply counts common
points of intersection of $A$, $B$, and $C$.  Since the two-point function
can also be expressed in terms of intersections, the Frobenius algebra
construction reproduces the familiar cup product on cohomology (which
counts intersections of the corresponding algebraic cycles), as
mentioned above.

For nonconstant maps, it turns out that the image of $\pi$ has a deformation
space whose dimension is  expected to  be $\dim X - 3$ (since the
spacetime has genus zero).\footnote{When the deformation space fails to have the expected
dimension, 
there is a natural way to integrate over the excess deformations
 to still produce a ``count'' of maps satisfying the three conditions
\cite{MOhep-th/9402147,MORuanTian,MOMR1467172}.}  
Imposing the conditions on $\pi(p)$, $\pi(q)$, and $\pi(r)$ then
cuts down the dimension to zero, and the corresponding maps can
be counted.  If we fix the homology class $\eta$ of the image,
then the Gromov--Witten invariant $GW^{X,\eta}_{0,3}(A,B,C)$,
which has a precise mathematical definition, is intended to count
the number of maps.

Each instanton contribution is weighted by the instanton
action $e^{\int_\eta S}= e^{2\pi i\tau \cdot \eta}$, which we often
denote by $q^\eta$.  The quantum product can be written as
\begin{equation}
A \star B := A \cup B 
+ \sum_{\eta} (A \star B)_\eta \, 
q^\eta
\label{MOeq:qproduct}
\end{equation}
where
$(A\star B)_\eta$ is the unique class satisfying
\begin{equation}
\left( (A\star B)_\eta \cup C \right)[X] =
GW^{X,\eta}_{0,3}(A,B,C) 
\end{equation}
for all $C$.
We can restrict the sum \eqref{MOeq:qproduct} to 
only range over those classes $\eta$ in $H_2(X,\mathbb{Z})$
whose intersection with each K\"ahler class is nonnegative.

Due to orbifold singularities in the deformation spaces,
the mathematical definition of  Gromov--Witten invariants only guarantees
that they are rational numbers, not integers.  The physical reason
for this is understood, and stems from a ``multiple covering'' phenomenon.
If the map $\pi$ factors through a multiple covering $S^2\to S^2$
of degree $m$, then the homology class of the image takes the form
$\eta=m \varphi$ but, as argued in \cite{MOtftrc} in dimension three
and \cite{MOhigherD} in arbitrary dimension (see also \cite{MOKlemm:2007in}),
the count of maps is the same.  

We can take this into account by defining a modified Gromov--Witten
invariant  $\widetilde{GW}^{X,\varphi}_{0,3}(A,B,C)$
which should only count the maps of degree one.  If we collect terms
according to degree one maps, we find a total instanton contribution of
\begin{equation}
\sum_{\varphi\in H_2(X,\mathbb{Z})} \widetilde{GW}^{X,\varphi}_{0,3}(A,B,C)
\frac{q^\varphi}{1-q^\varphi}
=
\sum_{\varphi\in H_2(X,\mathbb{Z})} \widetilde{GW}^{X,\varphi}_{0,3}(A,B,C)
\sum_{m=1}^\infty q^{m\varphi}
\ .
\end{equation}
Extracting the coefficient of $q^\eta$,  we obtain a formula
\begin{equation}
 {GW}^{X,\eta}_{0,3}(A,B,C)=
\sum_{\eta=m\varphi}
 \widetilde{GW}^{X,\varphi}_{0,3}(A,B,C)
\label{MOeq:modifiedGW}
\end{equation}
which can be used to define the modified Gromov--Witten invariants.  They
are expected to be integers.

For Calabi--Yau threefolds, there is one further modification which
can be made to the definitions.  The expected dimension of the deformation
spaces in that case is zero, so we expect only a finite number of possibilities
(in a fixed homology class) for the image of $\pi$.  For
each rational curve of class $\varphi$, there are $(A\cdot \varphi)$ choices for the location
of $\pi(p)$, $(B\cdot\varphi)$ choices for the location of $\pi(q)$,
and
$(C\cdot\varphi)$ choices for the  location of $\pi(r)$.  
It is then natural to define the ``Gromov--Witten instanton number''
of genus $0$ and class $\varphi$ to be
\begin{equation}
N_\varphi = \frac{\widetilde{GW}^{X,\varphi}_{0,3}(A,B,C)}{(A\cdot \varphi)
(B\cdot \varphi)(C\cdot \varphi)}
\label{MOeq:GWinstanton}
\end{equation}
which is expected to be an integer, independent of $A$, $B$, $C$, that
counts the number of rational curves in class $\varphi$.

Having spelled out in detail how instantons determine the quantum
cohomology ring, we can now explain the nonperturbative corrections
to the Zamoldchikov metric.  If we substitute
\begin{equation}
2\operatorname{Im}(t_j) = i(\bar{t}_j-t_j)
\end{equation}
into the perturbative expression for the metric \eqref{MOeq:perturbative-coho},
we obtain an expression involving  cup products
between twisted chiral operators (labeled by $t_j$) and their complex
conjugates (labeled by $\bar{t}_j$).  For the former, we can make
computations in the quantum cohomology ring instead, replacing $\cup$
by $\star$.  For the latter, we should use the complex-conjugated
cohomology ring, with instanton actions $\bar{q}^\eta$ rather than
$q^\eta$.  That is, we can do those computations in complex-conjugated
quantum cohomology, replacing $\cup$ by $\bar\star$.

In other words, the prescription for 
 nonperturbative corrections to the perturbative formula 
\eqref{MOeq:perturbative-coho} is: perform the multiplications
among holomorphic terms using $\star$ and among anti-holomorphic
terms using $\bar\star$; then multiply the pieces together using
cup product.\footnote{This prescription matches the formulas in
\cite{MO2-sphere} in dimension three and \cite{MOarXiv:1302.3760}
in dimension four, as well as considerations from mirror symmetry.}

Let us spell this out explicitly for Calabi--Yau threefolds. 
The perturbative expression can be written as
\begin{equation}
\frac i6 (\sum t_jL_j)^3 - \frac i2(\sum t_jL_j)^2(\sum \bar{t}_jL_j)
+\frac i2(\sum t_jL_j)(\sum \bar{t}_jL_j)^2 - \frac i6(\sum \bar{t}_jL_j)^3
-\frac{\zeta(3)}{4\pi^3}c_3(X).
\label{MOeq:4L-pert}
\end{equation}
When nonperturbative corrections are included, this becomes
\begin{equation}
\begin{aligned}
& \frac i6 \left(\sum t_jt_kt_\ell\left( (L_j {\cup} L_k {\cup} L_\ell)[X] +
\sum_\eta GW^{X,\eta}_{0,3}(L_j,L_k,L_\ell) q^\eta\right)\right) \\
&\quad - \frac i2\left(\sum t_jt_k\bar{t}_\ell\left( (L_j {\cup} L_k {\cup} L_\ell)[X] +
\sum_\eta GW^{X,\eta}_{0,3}(L_j,L_k,L_\ell) q^\eta\right)\right)\\
&\quad + \frac i2\left(\sum t_j\bar{t}_k\bar{t}_\ell\left( (L_j {\cup} L_k {\cup} L_\ell)[X] +
\sum_\eta GW^{X,\eta}_{0,3}(L_j,L_k,L_\ell) \bar{q}^\eta\right)\right) \\
&\quad -\frac i6 \left(\sum \bar{t}_j\bar{t}_k\bar{t}_\ell\left( (L_j {\cup} L_k {\cup} L_\ell)[X] +
\sum_\eta GW^{X,\eta}_{0,3}(L_j,L_k,L_\ell) \bar{q}^\eta\right)\right)
-\frac{\zeta(3)}{4\pi^3}c_3(X).
\end{aligned}
\label{MOeq:4L}
\end{equation}

\section{The two-sphere partition function and Gromov--Witten invariants}
\label{MOsec:S2}

\subsection{The $S^2$ partition function for a GLSM}
\label{MOsubsec:partition=G}

Consider an $\mathcal{N}=(2,2)$ GLSM with gauge group $G$,
chiral fields $\Phi_J$ of $R$-charge $q_J$
on which $G$ acts by the representation $\Psi:S\to SU(N)$,
superpotential
$W$, 
and complexified FI-parameters $\frac{\vartheta}{2\pi}+i\zeta
\in \operatorname{Ann}([\mathfrak{g},\mathfrak{g}])_{\mathbb{C}}
\subset \mathfrak{g}_{\mathbb{C}}^*$.
As in Section~\ref{MOsec:GLSM}, we fix a Cartan subgroup
$H$ of $G$ and weights $w_J\in \Lambda_{\text{wt}}\subset \mathfrak{h}^*$
describing the eigenvalues of $\Psi|_H$.\footnote{Note
that we are organizing things slightly differently from the way
the matter representation is described in \cite{MOBenini:2012ui,MODoroud:2012xw}.
With our conventions, each $\phi_J$ spans a one-dimensional space which is
preserved by the action of $H$ and thus is identified with a weight
of the representation $\Psi$.  In \cite{MOBenini:2012ui,MODoroud:2012xw},
the representation was decomposed into irreducible representations
of $G$  (labeled
by $\phi_J$, each of which had additional indices in those papers), and
then each of those irreducible representations
 was decomposed into weight spaces.}

The possible fluxes of the gauge theory through the $2$-sphere are GNO
quantized \cite{MOGoddard:1976qe}, which means that 
they are integer-valued functions
on the weight lattice, i.e., elements $\mathfrak{m}$ of the coweight
lattice $\Lambda_{\text{wt}}^\vee \subset\mathfrak{h}$.  
The combinations $i\sigma \pm \frac{\mathfrak{m}}2$ are what
appear in the formulas.

In \cite{MOBenini:2012ui,MODoroud:2012xw}, a computation of $Z_{S^2}$ 
for a $2$-sphere of radius $r$ is
made by expanding on the Coulomb branch and using localization,
after modifying the Lagrangian appropriately to put the theory
on $S^2$.
The original papers include twisted masses related to flavor symmetries
of the theory but we shall not include
those as they are not relevant for our application.
Since we are only studying the theories which are conformal in the
infrared, the only dependence on the radius is through a multiplicative
factor.
Some additional  notation: $|\mathcal{W}|$ denotes the order of 
the Weyl group of $G$, $\Delta^+$ denotes the set of positive
roots of $G$ (a subset of the weight lattice), and 
$\rho=\frac12\left(\sum_{\alpha\in\Delta^+}\alpha\right)$ is the Weyl vector.  
Here is the final formula, which assumes that all $R$-charges have been chosen
in the range $0<q<2$:

\begin{equation}
\frac{Z_{S^2}(z,\bar{z})}{(r/r_0)^{c/3}} 
= \frac{1}{|\mathcal{W}|} \sum_{\mathfrak{m}\in \Lambda_{\text{wt}}^\vee } 
\int_{\mathfrak{h}}  \left(\prod_{\mu=1}^{\operatorname{rank}(G)} \frac{d\sigma_{\mu}}{2\pi}\right)
  Z_{\text{class}} (\sigma, \mathfrak{m} )\ Z_{\text{gauge}} (\sigma, \mathfrak{m})\ 
Z_{\text{matter}}(\sigma,\mathfrak{m})
\, , \label{MOeq:ZS2-formula}
\end{equation}
where 
\begin{equation}
\begin{aligned}
Z_{\text{class}} &= \exp\langle \log z,i\sigma+\frac{\mathfrak{m}}2\rangle
\,\exp  \langle  \log \bar{z}, i\sigma - \frac{\mathfrak{m}}2\rangle 
= e^{-4\pi\langle\zeta,i\sigma\rangle+\langle i\vartheta,\mathfrak{m}\rangle}
\, \\
Z_{\text{gauge}} & = (-1)^{2 \rho\cdot \mathfrak{m}}\, \prod_{\alpha \in \Delta^+} 
\left( 
(\alpha\cdot \sigma)^2 
+
\frac14 (\alpha \cdot \mathfrak{m})^2 
\right)
 \ ,\\
Z_{\text{matter}} & = \prod_{J}
\frac{\Gamma\left(\frac{q_J}{2} -  w_J \cdot i\sigma 
- \frac{1}{2}w_J \cdot \mathfrak{m} 
\right)}
{\Gamma\left(1-\frac{q_J}{2}+w_J \cdot i\sigma
-\frac{1}{2}w_J\cdot \mathfrak{m} 
\right)}\ .
\end{aligned}
\end{equation}
We have included an overall sign in $Z_{\text{gauge}}$, missing
in \cite{MOBenini:2012ui,MODoroud:2012xw}, whose necessity was pointed out
in \cite{MOHonda:2013uca,MOHori:2013ika}.

\subsection{The hemisphere partition function and the  $tt^*$ equations}

As discussed above, the partition function of a GLSM has been
evaluated on a $2$-sphere \cite{MOBenini:2012ui,MODoroud:2012xw}.
The partition function has also been evaluated on a hemisphere, that is,
a half-sphere $D^2$ equipped with the spherical metric
\cite{MOSugishita:2013jca,MOHonda:2013uca,MOHori:2013ika},
as well as on real projective $2$-space 
\cite{MOKim:2013ola}.
A full discussion of these results is beyond the scope of this review,
but we will briefly present the result for the hemisphere, following
\cite{MOHori:2013ika}.

We need to specify BPS boundary conditions for the GLSM along the boundary
of the hemisphere, and the natural type of boundary conditions to use are
``B-branes''  
\cite{MOBecker:1995kb,MOOoguri:1996ck}.
The spectrum of B-branes is locally constant 
over $\mathcal{M}_{tc}$, and the data needed to specify B-branes
in a GLSM was thoroughly analyzed in \cite{MOhhp}.

Having specified some B-brane data $\mathcal{B}$, the partition
function $Z_{D^2,\mathcal{B}}$ on a hemisphere of radius $r$ is
evaluated explicitly in \cite{MOHori:2013ika}.  The dependence on
the radius is via an overall factor (which is the square root of
the corresponding factor in $Z_{S^2}$).
The dependence of the hemisphere partition function on the choice
$\mathcal{B}$ of  B-brane
comes through a factor $f_{\mathcal{B}}(\sigma)$ in
the integrand  described in \cite{MOHori:2013ika}, to
which we refer for further details (see also
\cite{MOKnapp:2016rec}).  

In order to evaluate the partition function, an appropriate integration
contour $\gamma\subset \mathfrak{h}_{\mathbb{C}}$ must be chosen.
With this understood, the result of \cite{MOHori:2013ika} is:
\begin{equation}
\frac{Z_{D^2,\mathcal{B}}(z,\bar{z})}{(r/r_0)^{c/6}} = 
\frac{1}{|\mathcal{W}|} 
\int_{\gamma}  
\prod_{\mu=1}^{\operatorname{rank}(G)} \frac{d\sigma_{\mu}}{2\pi}
\,e^{-4\pi i \langle \zeta,
 \sigma \rangle}
\prod_{\alpha \in \Delta^+} 
\alpha\cdot \sigma \operatorname{sinh}(\pi \alpha\cdot \sigma)
\prod_{J}
\Gamma\left(\frac{q_J}{2} -  w_J \cdot i\sigma 
\right)
f_{\mathcal{B}}(\sigma)
\, , \label{MOeq:ZD2-formula}
\end{equation}
where $\zeta = -\frac1{2\pi}\operatorname{Re}\log z$.
The authors of \cite{MOHori:2013ika} interpret this formula
as specifying a BPS charge for each choice of boundary condition,
and they verify that it agrees with the BPS charge in circumstances
under which both can be computed.

It is natural to expect that the hemisphere partition function
will play an important role in some yet-to-be-established holomorphic factorization
property for two-sphere partition functions.
Indeed, for the analogous three-dimensional gauge theories with $\mathcal{N}=2$ supersymmetry, the partition function computed in  \cite{MOKapustin:2009kz, MOHama:2011ea} displays such a factorization. (This fact was noted in \cite{MOPasquetti:2011fj}, and further explored in \cite{MOBeem:2012mb}.)
The authors of \cite{MOHori:2013ika} have taken an important step in
this direction in the 2D case by analyzing the GLSM partition function on an annulus
and studying how it can be used to glue together the results on the two
hemispheres to reproduce the result on the sphere.  Their calculation
is not completely general, but for theories with a geometric
phases they  verify that such a factorization result does indeed
hold.

An alternate approach to such a factorization result might proceed by
means of the ``$tt^*$ equations'' of \cite{MOCecotti:1991me},
as suggested in \cite{MO2-sphere,MOGomis:2012wy}.
The $tt^*$ equations describe the Zamolodchikov metric in terms
of topological twists of the GLSM: it is equal to the overlap of ground states in the $A$-twisted GLSM on one hemisphere, and an $\bar{A}$-twisted GLSM on the other hemisphere. 
The authors of \cite{MOGomis:2012wy} carry out a computation of such
an overlap by means of a calculation on the squashed two-sphere,
which has different limiting interpretations as the squashing parameter
is varied.  We refer to \cite{MOGomis:2012wy} for further details.

\subsection{Extracting Gromov--Witten invariants from the partition function}
\label{MOsec:procedure}

We now explain how, with the Euler characteristic $\chi(X)=c_3(X)$ as additional input, the relationship between the K\"ahler potential and the two-sphere
partition function  can be used to extract the Gromov--Witten invariants from the partition function $Z_{S^2}$ in the case of a Calabi--Yau threefold $X$.
The form of the partition function which we have determined for a nonlinear
sigma model depends on a choice of coordinates, so our task is to use
the asymptotic behavior of $Z_{S^2}$ to determine the appropriate coordinates.
For ease of exposition, in this section we will assume that we have chosen FI
coordinates so that a neighborhood of $\{z_a=0 \text{ for all } a\}$
describes a geometric phase of the GLSM.

To bring the partition function $Z_{S^2}(z,\bar{z})$ into an appropriate normal form and to extract the Gromov--Witten invariants, we use the following algorithm:

\begin{enumerate}
\item Evaluate $Z_{S^2}(z,\bar{z}) = e^{-K_{tc}}$ by contour integration as an expansion around large volume.
(We will discuss this step in more detail in the next section.)
The result can be expressed in terms of logarithmic coordinates
$\tau_j=\frac1{2\pi i}\log z_j$, and the goal is to find a change
of coordinates from $\tau_j$ to $t_j$.

\item Isolate the perturbative $\zeta(3)$ term and perform a K\"ahler transformation $K = K' + X^0(z) + \overline{X^0(z)}$ in order to reproduce the constant term $-\frac{\zeta(3)}{4\pi^3} \chi(X)$ in 
\eqref{MOeq:K4loop} and \eqref{MOeq:4L};
  (The initial $\zeta(3)$ term might
have a non-constant coefficient, which gives rise to a nontrivial
K\"ahler transformation.)

\item Read off the holomorphic part of the coefficient of $\bar{\tau}_j\bar{\tau}_k=\frac1{(2\pi i)^2}\log \bar{z}_j \log \bar{z}_k$, which should then be identified with
\begin{equation}
  -\frac{i}{2} \sum_\ell (L_j{\cup} L_k{\cup} L_\ell)[X] \, t_\ell \ .
\end{equation}
Use this to extract the NLSM coordinates $t_\ell$, which must have the form
\begin{equation} \label{MOeq:mmap}
  t_\ell = \frac{\log z_\ell}{2\pi i} + f_\ell(z)\ ,
\end{equation}
where $f_\ell(z)$ is a holomorphic function. 
This determines the NLSM coordinates up to the undetermined constants $f_\ell(0)$.
\item Invert the GLSM/NLSM map\footnote{This is the analogue of the
much-studied  ``mirror map'' relating chiral and twisted chiral conformal manifolds of a mirror pair \cite{MOmondiv}.} \eqref{MOeq:mmap} to obtain the $z_\ell$ as a function of $t_\ell$, 
\begin{equation}
z_\ell = e^{-2\pi i f_\ell(0)} \left( q_\ell + O(q^2)\right)\, ,
\end{equation}
where $q_\ell := e^{2\pi i t_\ell}$\,;

\item Fix the constant terms $f_\ell(0)$ by demanding the lowest order terms in the instanton expansion be positive; and, finally,

\item  Read off the (rational) Gromov--Witten invariants $GW^{X,\eta}_{0,3}(A,B,C)$
from the coefficients in the $q$-expansion. The (integral) Gromov--Witten instanton numbers $N_\eta$
of genus zero (roughly, the ``number of rational curves'') can then be obtained 
from the multi-covering formulas \eqref{MOeq:modifiedGW} and \eqref{MOeq:GWinstanton}.

\end{enumerate}

\section{Evaluating the partition function}
\label{MOsec:evaluating}

The low-energy effective theory describing the dynamics of the GLSM
depends on the value of the FI-parameters \cite{MOWitten:1993yc}. The space of FI-parameters can be divided into phase regions depending on the
character of the low-energy dynamics as explained in 
Section~\ref{MOsec:abelian-phases}.
In this section we show how this phase structure is closely
related to structure of the integrand of the two-sphere partition
function, and how this observation
can be used to determine Gromov--Witten invariants explicitly.

The idea is stated rather simply: when $Z_{S^2}$ is evaluated by the
method of residues, the contour prescription depends on the value of
the FI-parameters, which in turn affects the set of poles that
contribute to the integral. At certain codimension-one walls in 
FI-parameter space the structure of poles contributing to the $Z_{S^2}$
integral can change, signaling the presence of a GLSM phase transition
along that wall. In particular, for abelian GLSMs we show that this phase
structure is precisely the same secondary fan which governs
the low-energy physics.
Furthermore, we also describe how phases
of non-abelian GLSMs can be understood in terms of phases of an
associated ``Cartan'' theory.

\subsection{Analytic structure of the partition function}

When the
integrand in equation \eqref{MOeq:ZS2-formula} is analytically continued
to complex values of $\sigma\in \mathfrak{h}_{\mathbb{C}}$,
it becomes a meromorphic function of the integration variables. 
In order to evaluate this integral by means of residues, we need to
identify the location of all poles in that integrand.  We observe that
the gauge factor $Z_{\text{gauge}}$ never contributes poles and the integrand
has the same analytic structure if this term is omitted.  In fact,
if we retain the same matter content but restrict the gauge group
to a Cartan subgroup $H$, then up to a constant,
we simply omit the previous $Z_{\text{gauge}}$ factor
while retaining the $Z_{\text{classical}}$ and $Z_{\text{matter}}$ factors.
For the purposes of analyzing the analytic structure of the integrand,
we may thus restrict ourselves to abelian gauge theories without loss of generality.
See \cite{ContributionBL} for a further discussion of this point.

The partition function for an abelian GLSM with $G=H=U(1)^h$ is
\begin{equation}
\frac{Z_{S^2}(z,\bar{z})}{(r/r_0)^{c/3}} =  
\sum_{\mathfrak{m}\in \mathbb{Z}^h } 
\int_{\mathfrak{h}}  \left(\prod_{\mu=1}^{h} \frac{d\sigma_{\mu}}{2\pi}\right)
e^{-4\pi \langle \zeta, i\sigma \rangle + \langle i\vartheta,\mathfrak{m} \rangle}
\prod_{J}
\frac{\Gamma\left(\frac{q_J}{2} -  w_J \cdot i\sigma 
- \frac{1}{2}w_J \cdot \mathfrak{m} 
\right)}
{\Gamma\left(1-\frac{q_J}{2}+w_J \cdot i\sigma
-\frac{1}{2}w_J\cdot \mathfrak{m}
\right)}\ ,
\label{MOeq:ZS2-formula-abelian}
\end{equation}
assuming as in \eqref{MOeq:ZS2-formula} that $0<q_J<2$.
Recall that $\Gamma(z)$ is a meromorphic function in the complex plane with simple poles at $z = -n$, $n\in \mathbb{Z}_{\geq 0}$, with residue
$\operatorname{Res}_{z=-n} \Gamma(z) = \frac{(-1)^n}{n!}\,$ and an essential singularity at $z=\infty$. 
Taking into account cancellations between zeros and poles,
the $J^{\text{th}}$ factor in the final product of the integrand in 
\eqref{MOeq:ZS2-formula-abelian} has poles along the hyperplanes
\begin{equation}
  \label{MOeq:Zipoles}
H_J^{(k)}: \qquad  
\frac{q_J}{2} -  w_J \cdot i\sigma
- \frac{1}{2}w_J \cdot \mathfrak{m}
= -k\, , \quad k \in \mathbb{Z}_{\geq 0}, \quad k \geq 
w_J \cdot \mathfrak{m}\, .
\end{equation}
The partition function integrand can be regarded as a meromorphic function on the space $\mathbb{C}^{\operatorname{rank}(\mathfrak{g})}$ with poles along the hyperplanes $H_i^{(k)}$, which we refer to as \emph{polar divisors}. 
Note that the collection of hyperplanes $H_J^{(k)}$ is contained in the half space $\operatorname{Re}(w_J\cdot i\sigma)  \geq 0$, and also satisfies
$\operatorname{Im}(w_J\cdot i\sigma)=0$.
The analytic structure of the integrand will be relevant in what follows, as we will evaluate the integral in \eqref{MOeq:ZS2-formula-abelian} through the method of residues after choosing an appropriate way to close the integration
contour in $\mathfrak{h}_{\mathbb{C}}$.

\subsection{Residues and phases}
\label{MOsec:residues}

We use the multi-dimensional residue method \cite{MOGH,MOtsikh}  to evaluate the $h$-dimensional integral in \eqref{MOeq:ZS2-formula-abelian} 
for an abelian GLSM 
in terms of ``Grothendieck residues.'' 
The integration contour $\mathfrak{h}=\mathbb{R}^h$ must be replaced by
a closed contour $\gamma$ which meets none of the polar divisors.
This is done using a multi-dimensional analogue of the familiar Jordan lemma
which replaces an improper integral over the real axis by a contour
integral enclosing poles in the lower half-plane; the latter can
be evaluated using residues.
This can be done provided that the integrand
in the lower half-plane dies off at infinity in a suitable way.

Here, we do the same thing for each complex variable in 
$\mathfrak{h}_{\mathbb{C}}$.  The resulting integral is of a meromorphic
$h$-form over an $h$-dimensional compact cycle $\gamma$
which does not intersect
the poles of the integrand.  By Stokes' theorem, if we vary the
integration cycle without crossing any of the polar divisors, the value
of the integral does not change.  In particular, since a change of
basis of $\mathfrak{h}$ can be described by a path between the two
bases leading to a one-parameter family of contours $\gamma_t$;
as  long as none of the intermediate contours $\gamma_t$ intersect any of
the polar divisors, the integral does not change.  For our
primary integral \eqref{MOeq:ZS2-formula-abelian}, the growth
rate is controlled by the exponential factor and so it is the
sign of $\operatorname{Re}\langle \zeta, i\sigma\rangle$ which
matters; if that sign remains positive then the contours $\gamma_t$
will not encounter the polar divisors.

Each possible transverse intersection of $h$ polar divisors is associated to
an $h$-element subset $\mathcal{J}\subset\{1, \cdots, N\}$ 
such that the $h$ vectors $\{ {w}_J, {J \in \mathcal{J}} \}$ are linearly 
independent; as in Section~\ref{MOsec:abelian-phases},
we let $\mathfrak{J}$ denote the set of all such subsets $\mathcal{J}$. 
The corresponding polar divisors
$H^{(k)}_J$ for  all $J \in \mathcal{J}$ and $k \in \mathbb{Z}_{\geq 0}$ 
(defined in \eqref{MOeq:Zipoles}),
 intersect in an infinite discrete point set that we denote by $P_{\mathcal{J}}$. The residue of the integrand at any point $p \in P_{\mathcal{J}}$ is well defined and can be evaluated. 

To determine whether the residues at the points of $P_{\mathcal{J}}$
contribute to the integrand, we consider the basis $\{w_J, J\in\mathcal{J}\}$
of $\mathfrak{h}^*$, and 
express the FI-parameters
as  ${\zeta} = \sum_{J\in \mathcal{J}} \zeta_J w_J$ , (which
is possible since $\{w_J \ |\ J\in \mathcal{J}\}$ is linearly
independent). 
If $\zeta_J > 0$ for all $J \in \mathcal{J}$, then with respect
to the dual coordinates $\sigma_mu$ of $\mathfrak{h}$, the contours
all close in the lower half-plane and the residues at the points
of $P_{\mathcal{J}}$ are to be included in the integrand.  If
one of more $\zeta_J$ is negative, then at least one of the contours
closes in the upper half-plane rather than the lower half-plane,
and the residue is excluded.

Thus, the cones $\mathcal{C}_{\mathcal{J}}$ which determine the
phase structure of the theory (as explained in 
Section~\ref{MOsec:abelian-phases}) also determine which residues
to include in an evaluation of the two-sphere partition function 
\eqref{MOeq:ZS2-formula-abelian}.

Let $\mathcal{C}$ denote the (non-empty) intersection of all cones 
$C_{\mathcal{J}}$ that contain $\zeta$. The partition function, 
for ${\zeta}=-\frac1{2\pi}z \in \mathcal{C}$, can be evaluated as
\begin{equation}
\frac{Z_{S^2} (z,\bar{z})}{(r/r_0)^{c/3}} 
= \sum_{\begin{array}{c} \mathcal{J} \in \mathfrak{J}, \\ {\zeta} \in 
C_{\mathcal{J}}\end{array}} \sum_{p \in P_{\mathcal{J}}} \operatorname{Res}_{{\sigma} = {\sigma}_p} \left( \sum_{{m}\in \mathbb{Z}^s} \,\, 
e^{-4\pi\langle \zeta, i\sigma \rangle + \langle i\vartheta, \mathfrak{m} \rangle}\,\,
 \prod_{J=1}^N \,\, \frac{\Gamma(\frac{{q}_J}{2} - {w}_J \cdot({i\sigma} + \frac{{m}}{2}) )}{\Gamma(1-\frac{{q}_J}{2} + {w}_J\cdot ({i\sigma} - \frac{{m}}{2}) )} \right)\, ,
\end{equation}
where ${\sigma}_p$ denotes the coordinates of the point $p \in P_{\mathcal{J}}$. The above expression is an infinite series, whose convergence is controlled by the exponential factor 
$e^{-4\pi\langle \zeta, i\sigma \rangle + \langle i\vartheta, \mathfrak{m} \rangle}$,
up to a finite shift of  $i{\zeta}+\frac1{2\pi} {\vartheta}$ due to the exponential asymptotics of the remainder of the integrand. When ${\zeta}$ is sufficiently deep inside the cone $\mathcal{C}$ all summations are convergent. 

We thus see that the expression \eqref{MOeq:ZS2-formula-abelian} for $Z_{S^2}$ is an integral
of Mellin--Barnes type, which allows comparison of the behavior in different
regions of the FI-parameter space.    The Mellin--Barnes
technique had been used earlier to study GLSMs
\cite{MOPassare:1996db,MOmath.AG/9912109,MOMR2271990} but here it
arises as a property of the two-sphere partition function integrand,
rather than being introduced as a mathematical tool to aid in
understanding the theory.

\subsection[Gromov--Witten invariants of an example]{Gromov--Witten invariants of an example\footnote{I am grateful to my collaborators Jim Halverson, Hans Jockers, Vijay Kumar, Joshua Lapan, and Mauricio Romo for their assistance with this example.}} \label{MOsec:examples}

We consider again the example arising from the resolution of the degree eight hypersurface in the weighted projective space $\mathbb{P}^{(1,1,2,2,2)}$, using the GLSM description introduced in Section~\ref{MOsec:example}. The $R$-charges on the vector space $\mathbb{C}^7$ are specified by  the (rational) charge vector $(2-4{q}_1,{q}_2,{q}_2,{q}_1,{q}_1,{q}_1,{q}_1-2{q}_2)$ as a function of the two rational parameters ${q}_1$ and ${q}_2$. The requirement of all $R$-charges being positive gives the inequalities 
\begin{equation} 
0<{q}_2<\frac{{q}_1}2<\frac14 \ ,
\label{MO:positive-charge}
\end{equation}
and then it turns out that all $R$-charges are less than $2$ as well.
Deep in the geometric phyase (phase I),
we can simplify the expression
for the partition function \eqref{MOeq:ZS2-formula} by evaluating the $\sigma$-integration of the partition function with the help of residue calculus. 
Since all the K\"ahler parameters $\zeta_\ell$ are positive, we close the $\sigma$-integrations in the lower half-planes of the complexified $\sigma$-planes. 
The residues which contribute to the integral correspond to the
cones $C_{\mathcal{J}}$ illustrated in the first quadrant of Figure~\ref{MOfig:secondary}.
Computing the residues and bearing in mind
the charge condition \eqref{MO:positive-charge}, 
a straightforward but somewhat tedious algebraic manipulation
yields
the partition function 
\begin{equation}
\begin{aligned}
  Z_{S^2}(z,\bar{z}) & = |z_1|^{{q}_1}|z_2|^{{q}_2}\ \operatorname{Res}_{\vec\epsilon=0}\left[
  \frac{\pi^5 \sin 4\pi \epsilon_1}{\sin^3\pi\epsilon_1\cdot\sin^2\pi\epsilon_2\cdot\sin\pi(\epsilon_1-2\epsilon_2)}
  \right. \\
  &\ \times  \left(\sum_{k_1,k_2=0}^{+\infty} 
  \frac{\Gamma(4(k_1-\epsilon_1)+1)\,z_1^{k_1-\epsilon_1}z_2^{k_2-\epsilon_2}}{\Gamma((k_1-\epsilon_1)+1)^3\,\Gamma(k_2-\epsilon_2+1)^2\,\Gamma(k_1-\epsilon_1-2(k_2-\epsilon_2)+1)} \right) \\
  &\left.\ \times  \left(\sum_{k_1,k_2=0}^{+\infty} 
  \frac{\Gamma(4(k_1-\epsilon_1)+1)\,\bar z_1^{k_1-\epsilon_1}\bar z_2^{k_2-\epsilon_2}}{\Gamma((k_1-\epsilon_1)+1)^3\,\Gamma(k_2-\epsilon_2+1)^2\,\Gamma(k_1-\epsilon_1-2(k_2-\epsilon_2)+1)} \right)\right] \ .
\end{aligned}  
\end{equation}
We identify the partition function with the exponentiated K\"ahler potential, and we follow the algorithm in  Section~\ref{MOsec:procedure} to arrive at the K\"ahler potential in flat coordinates. Using the Euler characteristic $\chi=-168$ as an overall normalization for the $\zeta(3)$ term, we obtain the transformed K\"ahler potential $K'(z,\bar{z})$
\begin{equation}
  e^{-K'(z,\bar{z})} = -\frac1{8\pi^3|z_1|^{{q}_1}|z_2|^{{q}_2}} \frac{Z_{S^2}(z,\bar{z})}{|X^0(z)|^2}\ ,  ~~  
  X^0(z) = \sum_{k_1,k_2=0}^\infty  \frac{(4k_1)!}{(k_1!)^3\,(k_2!)^2\,(k_1-2k_2)!}z_1^{k_1}z_2^{k_2} \ .
\end{equation}
We observe that the relevant K\"ahler transformation involves the ``fundamental period'' $X^0(z)$ familiar from the toric mirror symmetry program \cite{MOBatyrev1993,MOBatyrev-Borisov-1994}. (This is a common feature of all of the examples which have been computed explicitly \cite{MO2-sphere}.) From the $(\log z_k\,\log z_\ell)$~terms, $k,\ell=1,2$, we extract the NLSM coordinates, which have the expansions
\begin{equation}
\begin{aligned}
  2\pi i\,t_1 & \,=\, \log z_1 + 104\,z_1-z_2+ 9\,780\,z_1^2+48\,z_1 z_2-\tfrac32\,z_2^2 + \ldots \ ,\\ 
  2\pi i\,t_2 & \,=\, \log z_2 +48\,z_1+2\,z_2  + 6\,408\,z_1^2-96\,z_1 z_2+3\,z_2^2+ \ldots \ .
\end{aligned}
\end{equation}
Inverting these maps, from the K\"ahler potential we find  the triple intersection numbers of the Calabi--Yau hypersurface
\begin{equation}
  L_1^3\,=\,8 \ , \qquad L_1^2L_2 \,=\,0 \ , \qquad L_1L_2^2\,=\, 4 \ , \qquad L_2^3\,=\,0 \ ,
\end{equation}
where $L_1$ and $L_2$ are the divisors associated to the K\"ahler coordinates $t_1$ and $t_2$,
and the genus zero Gromov--Witten instanton numbers $N_{d_1,d_2}$ of degree $(d_1,d_2)$ listed in Table~\ref{MOtab:Deg8GW}.
These same numbers had earlier been calculated by means of mirror symmetry 
\cite{MO2param1} but the present calculation is direct.

\begin{table}[t]
\begin{center}
\valign{\vfil#\vfil\cr
\footnotesize{
\vbox{\offinterlineskip
\halign{\MObstrut\vrule\ \hfil$#$\ \vrule\vrule&\ \hfil$#$\ &\ \hfil$#$\ &\ \hfil$#$\ &\ \hfil$#$\ &\ \hfil$#$\ &\ \hfil$#$\ &\ \hfil$#$\ \ \vrule\cr
\noalign{\hrule}
N_{d_1,d_2} & d_1\!=\!0 & 1 & 2 & 3 & 4 & 5 & 6 \cr
\noalign{\hrule\hrule}
d_2\!=\!0 & - & 640 & 10\,032 & 288\,384 & 10\,979\,984 & 495\,269\,504 & 24\,945\,542\,832 \cr 
1 &4 & 640 & 72\,224 & 7\,539\,200 & 757\,561\,520 & 74\,132\,328\,704 & 7\,117\,563\,990\,784 \cr 
2 &0 & 0 & 10\,032 & 7\,539\,200 & 2\,346\,819\,520 & 520\,834\,042\,880 & 95\,728\,361\,673\,920 \cr 
3 &0 & 0 & 0 & 288\,384 & 757\,561\,520 & 520\,834\,042\,880 & 212\,132\,862\,927\,264 \cr 
4 &0 & 0 & 0 & 0 & 10\,979\,984 & 74\,132\,328\,704 & 95\,728\,361\,673\,920 \cr 
5 &0 & 0 & 0 & 0 & 0 & 495\,269\,504 & 7\,117\,563\,990\,784 \cr 
6 &0 & 0 & 0 & 0 & 0 & 0 & 24\,945\,542\,832 \cr 
7 &0 & 0 & 0 & 0 & 0 & 0 & 0 \cr 
\noalign{\hrule}}
}}\cr}
\centering
\caption{Gromov--Witten instanton numbers of the blown up degree eight Calabi--Yau hypersurface  in $\widetilde{\mathbb{P}^{(1,1,2,2,2)}}$ (genus zero).}
\label{MOtab:Deg8GW}
\end{center}
\end{table}

Our example Calabi--Yau threefold $X$ exhibits an extremal transition to a different Calabi--Yau threefold $\widehat X$ \cite{MO2param1}. Such extremal transitions have been studied in detail in \cite{MOclemens1983double,MOMR848512,MOMR909231,MOCandelas:1989ug,MObhole}, and a transition arises for the given example as follows. As the curves in the homology class dual to the divisor $L_2$ in $X$ are blown down, $X$ develops (generically) four nodal singularities. The resulting singular threefold $X_\text{sing}$ is a degree eight hypersurface in $\mathbb{P}^{(1,1,2,2,2)}$ with the nodal points induced from the singularities of the weighted projective space. Alternatively, we may embed $X_\text{sing}$ into $\mathbb{P}^5$ as the complete intersection of the degree two polynomial $\widehat F_{2}(\eta_1,\ldots,\eta_6)=\eta_1\eta_2-\eta_3^2$ and a degree four polynomial $\widehat F_{4}(\eta_1,\ldots,\eta_6)$ with the homogeneous coordinates $\eta_1,\ldots,\eta_6$ of $\mathbb{P}^5$. $X_\text{sing}$ embedded in $\mathbb{P}^{(1,1,2,2,2)}$ is associated to the degree eight polynomial $F_{8}(\phi_1,\ldots,\phi_5)=\widehat F_{4}(\phi_1^2,\phi_2^2,\phi_1\phi_2,\phi_3,\phi_4,\phi_5)$ in terms of the weighted homogeneous coordinates $\phi_1,\ldots,\phi_5$ of $\mathbb{P}^{(1,1,2,2,2)}$. Perturbing the polynomial $\widehat F_{2}$, we obtain the smooth and deformed Calabi--Yau threefold $\widehat X$ as the complete intersection
of degree $(2,4)$ in $\mathbb{P}^5$.

Due to this extremal transition, the genus zero Gromov--Witten instanton numbers of $\widehat X$ appear as the sum \cite{MOMR1839289}
\begin{equation}
  N_{\delta}(\widehat X)\,=\, \sum_{d=0}^{+\infty} N_{\delta,d}( X) \ .
\end{equation}
From Table~\ref{MOtab:Deg8GW}, we extract the invariants
\begin{equation}
   N_\delta(\widehat X)=1\,280\, ,\ 92\,288 \, ,\ 15\,655\,168 \, ,\ 3\,883\,902\,528 \, ,\ 1\,190\,923\,282\,176\, ,\ldots \ ,
\end{equation}
which are in agreement with the genus zero Gromov--Witten instanton numbers of $\mathbb{P}^5[2,4]$ calculated (again using mirror symmetry) in~\cite{MOMR1201748}.

\bigskip

\subsection*{Acknowledgments}

I would like to thank Jim Halverson, Hans Jockers,
Vijay Kumar, Josh Lapan, 
Ronen Plesser,
and Mauricio Romo for collaboration on
various of the results presented here.
I particularly thank 
Nati Seiberg for sharing insights about 
his
work, some of which is also presented here.
I also thank
Chris Beem,
Francesco Benini,
Eduardo Cattani,
Cyril Closset, 
Jacques Distler,
Jaume Gomis,
Rajesh Gopakumar, 
Antonella Grassi, 
Simeon Hellerman,
Kentaro Hori,
Shinobu Hosono, 
Nabil Iqbal,
Zohar Komargodsky,
Peter Koroteev,
Sungjay Lee,
Bruno Le~Floch, 
Jan Manschot, 
Greg Moore, 
Joe Polchinski,
Eric Sharpe,
Samson Shatashvili,
Yuji Tachikawa,
Cumrun Vafa,
and
Edward Witten
for useful discussions and correspondence.
I would also like to thank the Institut Henri Poincar\'e for hospitality
during the final stages of this project.
This work was supported in part by National Science Foundation 
Grant PHY-1307513 (USA) and by the Centre National de la Recherche
Scientifique (France).

\bigskip

\section{Appendix. $\mathcal{N}=(2,2)$ supersymmetry in two dimensions}
\label{MOapp:N22}

We study two-dimensional theories with $\mathcal{N}=(2,2)$ supersymmetry.
There are two basic types of supermultiplet, the notation for which is
obtained by dimensional reduction from that of \cite{MOMR701704}.
A {\em chiral supermultiplet}\/ has components $(\phi,\psi,F)$
while a {\em vector supermultiplet}\/ has components $(v,\sigma,\lambda, D)$.

In Euclidean signature, we use a complex coordinate $z$ on
spacetime, and consider $\sqrt{dz}$ and $\sqrt{d\bar z}$
as bases for the two spinor bundles $S_+$ and $S_-$ of opposite chirality.
In components, we can write
\begin{equation}
\psi = \psi_- \sqrt{dz} + \psi_+ \sqrt{d\bar z}.
\end{equation}

In Minkowski signature, with time coordinate $x^0$ and spatial coordinate
$x^1$, the metric has components $\eta_{00}=-1$, $\eta_{11}=1$, $\eta_{01}=0$.
The
fermionic coordinates
$\theta^\pm$ and $\bar \theta^\pm$ are complex, related by complex 
conjugation (i.e., $(\theta^\pm)^* = \bar\theta^\pm$).  The $\pm$
indices denote chirality.  
In particular, 
\begin{equation}\begin{bmatrix} \cosh\gamma&\sinh\gamma\\ \sinh\gamma&\cosh\gamma \end{bmatrix} \in SO(1,1)\end{equation}
acts by $e^{\pm\gamma/2}$ on $\theta^{\pm}$ and
also by $e^{\pm\gamma/2}$ on $\bar\theta^{\pm}$.

A superfield is a function $\Phi$ of these variables, and can be expanded
into the thetas.  $\Phi$ is bosonic if $[\theta^\alpha,\Phi]=0$
and fermionic if $\{\theta^\alpha,\Phi\}=0$.

We introduce the combinations $x^\pm := x^0 \pm x^1$ and the
corresponding differential operators 
\begin{equation}
\partial_\pm = \frac\partial{\partial x^\pm}
:= \frac12\left(\frac\partial{\partial x^0} \pm \frac\partial{\partial x^1}
\right).
\end{equation}

There are two natural sets of differential operators on superspace:
\begin{align}
\mathcal{Q}^\pm  =& \frac\partial{\partial \theta^\pm} + i\bar\theta^\pm 
\partial_\pm\\
\overline{\mathcal{Q}}^\pm  =& -\frac\partial{\partial \bar\theta^\pm} - i\theta^\pm 
\partial_\pm
\end{align}
which satisfy 
$\{ \mathcal{Q}_\pm, \overline{\mathcal{Q}}_\pm\} =-2i\partial_\pm$, and
\begin{align}
D^\pm  =& \frac\partial{\partial \theta^\pm} - i\bar\theta^\pm 
\partial_\pm\\
\overline{D}^\pm  =& -\frac\partial{\partial \bar\theta^\pm} + i\theta^\pm 
\partial_\pm
\end{align}
which anti-commute with the first set, and satisfy
$\{ D_\pm, \overline{D}_\pm\} =2i\partial_\pm$.

A {\em chiral superfield}\/ $\Phi$ satisfies $\overline{D}_\pm\Phi=0$.
Its component description is:
\begin{equation}
\Phi(x^\mu,\theta^\pm,\bar\theta^\pm)= \phi(y^\pm)
+\theta^\alpha \psi_\alpha(y^\pm) + \theta^+\theta^-F(y^\pm)
\end{equation}
where $y^\pm= x^\pm -i\theta^\pm\bar\theta^\pm$
and $F(y^\pm)$ is a non-propagating ``auxiliary'' field in the multiplet..
A {\em twisted chiral superfield}\/ $U$ satisfies 
$\overline{D}_+U = D_-U = 0$.

A vector multiplet is a real superfield $V=V(x^\mu,\theta^\pm,\bar\theta^\pm)$
which under a gauge transformation $\Phi\mapsto e^{iA}\Phi$ transforms
as $V\mapsto V + i(\overline{A}-A)$, so that the Lagrangian
$$ \int d^4\theta \bar\Phi e^V \Phi$$
is invariant under gauge transformations.

There is a gauge transformation putting the gauge field into
{\em Wess-Zumino gauge}, where it takes the form
\begin{multline}
 V = \theta^-\bar\theta^-(v_0-v_1)
+ \theta^+\bar\theta^+(v_0+v_1)
-\theta^-\bar\theta^+\sigma
-\theta^+\bar\theta^-\bar\sigma
+i\theta^-\theta^+(\bar\theta^-\bar\lambda_-+\bar\theta^+\bar\lambda_+)\\
+i\bar\theta^+\bar\theta^-(\theta^-\lambda_-+\theta^+\lambda_+)+
\theta^-\theta^+\bar\theta^+\bar\theta^-D
\end{multline}
where $\sigma$ is a complex scalar field, $\lambda_\pm$ and $\bar\lambda_\pm$
define a Dirac fermion, and $D$ is an auxiliary real scalar field.
The components $v_\mu$ define the covariant derivatives
$D_\mu = \partial_\mu + iv_\mu$.

The field strength of $V$ is
\begin{equation}
 \Sigma :=\overline{D}_+ D_- V
\end{equation}
and it has a component expansion
\begin{equation}
 \Sigma = \sigma(\tilde y)
+ i\theta^+\bar\lambda_+(\tilde{y})-i\bar\theta^-\lambda_-(\tilde{y})
+\theta^+\bar\theta^-[D(\tilde y)-iv_{01}(\tilde y)], 
\end{equation}
where $\tilde y^\pm :=x^\pm \mp i\theta^\pm\bar\theta^\pm$
and $v_{01}=\partial_0v_1-\partial_1v_0$ is the curvature of the connection.

\documentfinish
\begin{thebibliography}{100}

\bibitem{ContributionSummary}
V.~Pestun and M.~Zabzine, eds., {\em Localization techniques in quantum field
  theory}, vol.~xx.
\newblock Journal of Physics A, 2016.
\newblock \href{http://arxiv.org/abs/1608.02952}{{\tt 1608.02952}}.
\newblock \url{https://arxiv.org/src/1608.02952/anc/LocQFT.pdf},
  \url{http://pestun.ihes.fr/pages/LocalizationReview/LocQFT.pdf}.

\bibitem{MOGreene:1990ud}
B.~R. Greene and M.~R. Plesser, ``Duality in {C}alabi--{Y}au Moduli Space,''
{\em Nucl. Phys. B} {\bf 338} (1990)  15--37.
%%CITATION = NUPHA,B338,15;%%.

\bibitem{MOCLS}
P.~Candelas, M.~Lynker, and R.~Schimmrigk, ``{C}alabi--{Y}au Manifolds in
  Weighted {$\mathbb P_4$},''
{\em Nucl. Phys. B} {\bf 341} (1990)  383--402.
%%CITATION = NUPHA,B341,383;%%.

\bibitem{MOAspinwall:1990xe}
P.~S. Aspinwall, C.~A. L\"utken, and G.~G. Ross, ``Construction and Couplings
  of Mirror Manifolds,''
{\em Phys. Lett. B} {\bf 241} (1990)  373--380.
%%CITATION = PHLTA,B241,373;%%.

\bibitem{MOCDGP}
P.~Candelas, X.~C. de~la Ossa, P.~S. Green, and L.~Parkes, ``A Pair of
  {C}alabi--{Y}au Manifolds as an Exactly Soluble Superconformal Theory,''
{\em Nucl. Phys. B} {\bf 359} (1991)  21--74.
%%CITATION = NUPHA,B359,21;%%.

\bibitem{MOGromov}
M.~Gromov, ``Pseudo Holomorphic Curves in Symplectic Manifolds,'' {\em Invent.
  Math.} {\bf 82} (1985)  307--347.

\bibitem{MOtsm}
E.~Witten, ``Topological Sigma Models,'' {\em Commun. Math. Phys.} {\bf 118}
  (1988)  411--449.

\bibitem{MOcomplete1}
A.~B. Givental, ``Equivariant {G}romov--{W}itten invariants,'' {\em Internat.
  Math. Res. Notices} (1996) no.~13, 613--663,
  \href{http://arxiv.org/abs/arXiv:alg-geom/9603021}{{\tt
  arXiv:alg-geom/9603021}}.

\bibitem{MOcomplete2}
B.~H. Lian, K.~Liu, and S.-T. Yau, ``Mirror Principle. {I},'' {\em Asian J.
  Math.} {\bf 1} (1997) no.~4, 729--763,
\href{http://arxiv.org/abs/arXiv:alg-geom/9712011}{{\tt
  arXiv:alg-geom/9712011}}.
%%CITATION = 00204,1,729;%%.

\bibitem{MOarXiv:1201.6350}
Y.~Cooper and A.~Zinger, ``Mirror Symmetry for Stable Quotients Invariants,''
  {\em Michigan Math. J.} {\bf 63} (2014) no.~3, 571--621,
  \href{http://arxiv.org/abs/arXiv:1201.6350 [math.AG]}{{\tt arXiv:1201.6350
  [math.AG]}}.

\bibitem{MO2-sphere}
H.~Jockers, V.~Kumar, J.~M. Lapan, D.~R. Morrison, and M.~Romo, ``Two-Sphere
  Partition Functions and {G}romov--{W}itten Invariants,'' {\em Commun. Math.
  Phys.} {\bf 325} (2014)  1139--1170,
\href{http://arxiv.org/abs/arXiv:1208.6244 [hep-th]}{{\tt arXiv:1208.6244
  [hep-th]}}.
%%CITATION = ARXIV:1208.6244;%%.

\bibitem{MOPestun:2007rz}
V.~Pestun, ``Localization of Gauge Theory on a Four-Sphere and Supersymmetric
  {W}ilson Loops,'' {\em Commun. Math. Phys.} {\bf 313} (2012)  71--129,
\href{http://arxiv.org/abs/0712.2824}{{\tt arXiv:0712.2824 [hep-th]}}.
%%CITATION = ARXIV:0712.2824;%%.

\bibitem{MOBenini:2012ui}
F.~Benini and S.~Cremonesi, ``Partition Functions of {$\mathcal{N}=(2,2)$}
  Gauge Theories on {S$^{2}$} and Vortices,'' {\em Commun. Math. Phys.} {\bf
  334} (2015) no.~3, 1483--1527,
\href{http://arxiv.org/abs/1206.2356}{{\tt arXiv:1206.2356 [hep-th]}}.
%%CITATION = ARXIV:1206.2356;%%.

\bibitem{MODoroud:2012xw}
N.~Doroud, J.~Gomis, B.~Le~Floch, and S.~Lee, ``Exact Results in {D=2}
  Supersymmetric Gauge Theories,'' {\em JHEP} {\bf 1305} (2013)  093,
\href{http://arxiv.org/abs/1206.2606}{{\tt arXiv:1206.2606 [hep-th]}}.
%%CITATION = ARXIV:1206.2606;%%.

\bibitem{MOGomis:2012wy}
J.~Gomis and S.~Lee, ``Exact {K}{\"a}hler Potential from Gauge Theory and
  Mirror Symmetry,'' {\em JHEP} {\bf 1304} (2013)  019,
\href{http://arxiv.org/abs/1210.6022}{{\tt arXiv:1210.6022 [hep-th]}}.
%%CITATION = ARXIV:1210.6022;%%.

\bibitem{MOGerchkovitz:2014gta}
E.~Gerchkovitz, J.~Gomis, and Z.~Komargodski, ``Sphere Partition Functions and
  the {Z}amolodchikov Metric,'' {\em JHEP} {\bf 1411} (2014)  001,
\href{http://arxiv.org/abs/1405.7271}{{\tt arXiv:1405.7271 [hep-th]}}.
%%CITATION = ARXIV:1405.7271;%%.

\bibitem{MOGomis:2015yaa}
J.~Gomis, P.-S. Hsin, Z.~Komargodski, A.~Schwimmer, N.~Seiberg, and S.~Theisen,
  ``Anomalies, Conformal Manifolds, and Spheres,'' {\em JHEP} {\bf 1603} (2016)
   022,
\href{http://arxiv.org/abs/1509.08511}{{\tt arXiv:1509.08511 [hep-th]}}.
%%CITATION = ARXIV:1509.08511;%%.

\bibitem{MOWitten:1993yc}
E.~Witten, ``Phases of {$N = 2$} Theories in Two Dimensions,'' {\em Nucl. Phys.
  B} {\bf 403} (1993)  159--222,
\href{http://arxiv.org/abs/arXiv:hep-th/9301042}{{\tt arXiv:hep-th/9301042}}.
%%CITATION = HEP-TH 9301042;%%.

\bibitem{MOsumming}
D.~R. Morrison and M.~R. Plesser, ``Summing the Instantons: Quantum Cohomology
  and Mirror Symmetry in Toric Varieties,'' {\em Nucl. Phys. B} {\bf 440}
  (1995)  279--354,
\href{http://arxiv.org/abs/arXiv:hep-th/9412236}{{\tt arXiv:hep-th/9412236}}.
%%CITATION = HEP-TH 9412236;%%.

\bibitem{MOtowards-duality}
D.~R. Morrison and M.~R. Plesser, ``Towards Mirror Symmetry as Duality for
  Two-Dimensional Abelian Gauge Theories,'' in {\em Trieste Conference on
  S-Duality and Mirror Symmetry}, vol.~46 of {\em Nucl. Phys. B Proc. Suppl.},
  pp.~177--186.
\newblock 1996.
\newblock
\href{http://arxiv.org/abs/arXiv:hep-th/9508107}{{\tt arXiv:hep-th/9508107}}.
\newblock
%%CITATION = HEP-TH 9508107;%%.

\bibitem{MOnew-methods}
J.~Halverson, V.~Kumar, and D.~R. Morrison, ``New Methods for Characterizing
  Phases of {2D} Supersymmetric Gauge Theories,'' {\em JHEP} {\bf 1309} (2013)
  143,
\href{http://arxiv.org/abs/arXiv:1305.3278 [hep-th]}{{\tt arXiv:1305.3278
  [hep-th]}}.
%%CITATION = ARXIV:1305.3278;%%.

\bibitem{MOgamma-class}
J.~Halverson, H.~Jockers, J.~M. Lapan, and D.~R. Morrison, ``Perturbative
  Corrections to {K}{\"a}hler Moduli Spaces,'' {\em Commun. Math. Phys.} {\bf
  333} (2015)  1563--1584,
\href{http://arxiv.org/abs/arXiv:1308.2157 [hep-th]}{{\tt arXiv:1308.2157
  [hep-th]}}.
%%CITATION = ARXIV:1308.2157;%%.

\bibitem{MOZamolodchikov:1986gt}
A.~Zamolodchikov, ``Irreversibility of the Flux of the Renormalization Group in
  a {2D} Field Theory,''
{\em JETP Lett.} {\bf 43} (1986)  730--732.
%%CITATION = JTPLA,43,730;%%.

\bibitem{MOSeiberg:1993vc}
N.~Seiberg, ``Naturalness Versus Supersymmetric Nonrenormalization Theorems,''
  {\em Phys. Lett. B} {\bf 318} (1993)  469--475,
\href{http://arxiv.org/abs/arXiv:hep-ph/9309335}{{\tt arXiv:hep-ph/9309335}}.
%%CITATION = HEP-PH/9309335;%%.

\bibitem{MOOsborn:1991gm}
H.~Osborn, ``Weyl Consistency Conditions and a Local Renormalization Group
  Equation for General Renormalizable Field Theories,''
{\em Nucl. Phys. B} {\bf 363} (1991)  486--526.
%%CITATION = NUPHA,B363,486;%%.

\bibitem{MOFestuccia:2011ws}
G.~Festuccia and N.~Seiberg, ``Rigid Supersymmetric Theories in Curved
  Superspace,'' {\em JHEP} {\bf 1106} (2011)  114,
\href{http://arxiv.org/abs/1105.0689}{{\tt arXiv:1105.0689 [hep-th]}}.
%%CITATION = ARXIV:1105.0689;%%.

\bibitem{MOClosset:2014pda}
C.~Closset and S.~Cremonesi, ``Comments on {$ \mathcal{N} $} = (2, 2)
  Supersymmetry on Two-Manifolds,'' {\em JHEP} {\bf 1407} (2014)  075,
\href{http://arxiv.org/abs/1404.2636}{{\tt arXiv:1404.2636 [hep-th]}}.
%%CITATION = ARXIV:1404.2636;%%.

\bibitem{MOBae:2015eoa}
J.~Bae, C.~Imbimbo, S.-J. Rey, and D.~Rosa, ``{New Supersymmetric Localizations
  from Topological Gravity},''
  \href{http://dx.doi.org/10.1007/JHEP03(2016)169}{{\em JHEP} {\bf 03} (2016)
  169},
\href{http://arxiv.org/abs/1510.00006}{{\tt arXiv:1510.00006 [hep-th]}}.
%%CITATION = ARXIV:1510.00006;%%.

\bibitem{MOHowe:1987ba}
P.~S. Howe and G.~Papadopoulos, ``{N=2}, {D = 2} Supergeometry,''
{\em Class. Quant. Grav.} {\bf 4} (1987)  11--21.
%%CITATION = CQGRD,4,11;%%.

\bibitem{MOAdams:2011vw}
A.~Adams, H.~Jockers, V.~Kumar, and J.~M. Lapan, ``{N=1} Sigma Models in
  {$AdS_4$},'' {\em JHEP} {\bf 1112} (2011)  042,
\href{http://arxiv.org/abs/1104.3155}{{\tt arXiv:1104.3155 [hep-th]}}.
%%CITATION = ARXIV:1104.3155;%%.

\bibitem{MOBershadsky:1995sp}
M.~Bershadsky, C.~Vafa, and V.~Sadov, ``D-Strings on {D}-Manifolds,'' {\em
  Nucl. Phys. B} {\bf 463} (1996)  398--414,
\href{http://arxiv.org/abs/hep-th/9510225}{{\tt hep-th/9510225}}.
%%CITATION = HEP-TH 9510225;%%.

\bibitem{MOenhanced}
S.~Katz, D.~R. Morrison, and M.~R. Plesser, ``Enhanced Gauge Symmetry in Type
  {II} String Theory,'' {\em Nucl. Phys. B} {\bf 477} (1996)  105--140,
\href{http://arxiv.org/abs/arXiv:hep-th/9601108}{{\tt arXiv:hep-th/9601108}}.
%%CITATION = HEP-TH 9601108;%%.

\bibitem{MOGrisaru:1986dk}
M.~T. Grisaru, A.~E.~M. van~de Ven, and D.~Zanon, ``Two-Dimensional
  Supersymmetric Sigma-Models on {R}icci-Flat {K}\"ahler Manifolds are Not
  Finite,''
{\em Nucl. Phys. B} {\bf 277} (1986) no.~2, 388--408.
%%CITATION = NUPHA,B277,388;%%.

\bibitem{MONemeschansky:1986yx}
D.~Nemeschansky and A.~Sen, ``Conformal Invariance of Supersymmetric
  {$\sigma$}-Models on {C}alabi-{Y}au Manifolds,''
{\em Phys. Lett. B} {\bf 178} (1986) no.~4, 365--369.
%%CITATION = PHLTA,B178,365;%%.

\bibitem{MOCandelas:1989qn}
P.~Candelas, T.~Hubsch, and R.~Schimmrigk, ``Relation Between the
  {W}eil-Petersson and {Z}amolodchikov Metrics,''
{\em Nucl. Phys. B} {\bf 329} (1990)  583--590.
%%CITATION = NUPHA,B329,583;%%.

\bibitem{MOTian}
G.~Tian, ``Smoothness of the Universal Deformation Space of Compact
  {C}alabi--{Y}au Manifolds and its {P}eterson--{W}eil Metric,'' in {\em
  Mathematical Aspects of String Theory}, S.-T. Yau, ed., pp.~629--646.
\newblock World Scientific, Singapore, 1987.

\bibitem{MOTodorov-weil-petersson}
A.~N. Todorov, ``The {W}eil-{P}etersson Geometry of the Moduli Space of
  {$SU(n{\ge}3)$} ({C}alabi--{Y}au) Manifolds, {I},'' {\em Commun. Math. Phys.}
  {\bf 126} (1989)  325--246.

\bibitem{MOGrisaru:1979wc}
M.~T. Grisaru, W.~Siegel, and M.~Rocek, ``Improved Methods for Supergraphs,''
{\em Nucl. Phys. B} {\bf 159} (1979)  429--250.
%%CITATION = NUPHA,B159,429;%%.

\bibitem{MOcompact}
D.~R. Morrison, ``Compactifications of Moduli Spaces Inspired by Mirror
  Symmetry,'' in {\em Journ\'ees de G\'eom\'etrie Alg\'ebrique d'Orsay (Juillet
  1992)}, vol.~218 of {\em Ast\'erisque}, pp.~243--271.
\newblock Soci\'et\'e Math\-\'ematique de France, 1993.
\newblock \href{http://arxiv.org/abs/arXiv:alg-geom/9304007}{{\tt
  arXiv:alg-geom/9304007}}.

\bibitem{MOHori:2011pd}
K.~Hori, ``Duality In Two-Dimensional {(2,2)} Supersymmetric Non-{A}belian
  Gauge Theories,'' {\em JHEP} {\bf 10} (2013)  121,
\href{http://arxiv.org/abs/arXiv:1104.2853 [hep-th]}{{\tt arXiv:1104.2853
  [hep-th]}}.
%%CITATION = 1104.2853;%%.

\bibitem{MOHori:2013gga}
K.~Hori and J.~Knapp, ``Linear Sigma Models With Strongly Coupled Phases -- One
  Parameter Models,'' {\em JHEP} {\bf 11} (2013)  070,
\href{http://arxiv.org/abs/1308.6265}{{\tt arXiv:1308.6265 [hep-th]}}.
%%CITATION = ARXIV:1308.6265;%%.

\bibitem{MOKnapp:2016rec}
J.~Knapp, M.~Romo, and E.~Scheidegger, ``Hemisphere Partition Function and
  Analytic Continuation to the Conifold Point,''
\href{http://arxiv.org/abs/1602.01382}{{\tt arXiv:1602.01382 [hep-th]}}.
%%CITATION = ARXIV:1602.01382;%%.

\bibitem{MOmondiv}
P.~S. Aspinwall, B.~R. Greene, and D.~R. Morrison, ``The Monomial-Divisor
  Mirror Map,'' {\em Internat. Math. Res. Notices} (1993)  319--337,
  \href{http://arxiv.org/abs/arXiv:alg-geom/9309007}{{\tt
  arXiv:alg-geom/9309007}}.

\bibitem{MOAspinwall:2015zia}
P.~S. Aspinwall and M.~R. Plesser, ``General Mirror Pairs for Gauged Linear
  Sigma Models,'' {\em JHEP} {\bf 11} (2015)  029,
\href{http://arxiv.org/abs/1507.00301}{{\tt arXiv:1507.00301 [hep-th]}}.
%%CITATION = ARXIV:1507.00301;%%.

\bibitem{MOSilverstein:1994ih}
E.~Silverstein and E.~Witten, ``Global {$U(1)$} {R} Symmetry and Conformal
  Invariance of (0,2) Models,'' {\em Phys. Lett. B} {\bf 328} (1994)  307--311,
\href{http://arxiv.org/abs/arXiv:hep-th/9403054}{{\tt arXiv:hep-th/9403054}}.
%%CITATION = HEP-TH 9403054;%%.

\bibitem{MOMR1020882}
I.~M. Gel$'$fand, A.~V. Zelevinski{\u\i}, and M.~M. Kapranov, ``Newton
  Polyhedra of Principal {$A$}-Determinants,'' {\em Soviet Math. Dokl.} {\bf
  40} (1990)  278--281.

\bibitem{MOMR1073208}
I.~M. Gel$'$fand, A.~V. Zelevinski{\u\i}, and M.~M. Kapranov, ``Discriminants
  of Polynomials in Several Variables and Triangulations of {N}ewton
  Polyhedra,'' {\em Leningrad Math.\ J.} {\bf 2} (1991)  449--505.

\bibitem{MOBFS}
L.~J. Billera, P.~Filliman, and B.~Sturmfels, ``Constructions and Complexity of
  Secondary Polytopes,'' {\em Adv. Math.} {\bf 83} (1990) no.~2, 155--179.

\bibitem{MOcatp}
P.~S. Aspinwall, B.~R. Greene, and D.~R. Morrison, ``Calabi--{Y}au Moduli
  Space, Mirror Manifolds and Spacetime Topology Change in String Theory,''
  {\em Nucl. Phys. B} {\bf 416} (1994)  414--480,
\href{http://arxiv.org/abs/arXiv:hep-th/9309097}{{\tt arXiv:hep-th/9309097}}.
%%CITATION = HEP-TH 9309097;%%.

\bibitem{MO2param1}
P.~Candelas, X.~de~la Ossa, A.~Font, S.~Katz, and D.~R. Morrison, ``Mirror
  Symmetry for Two Parameter Models -- {I},'' {\em Nucl. Phys. B} {\bf 416}
  (1994)  481--562,
\href{http://arxiv.org/abs/arXiv:hep-th/9308083}{{\tt arXiv:hep-th/9308083}}.
%%CITATION = HEP-TH 9308083;%%.

\bibitem{MOHosono:1993qy}
S.~Hosono, A.~Klemm, S.~Theisen, and S.-T. Yau, ``Mirror Symmetry, Mirror Map
  and Applications to {C}alabi--{Y}au Hypersurfaces,'' {\em Commun. Math.
  Phys.} {\bf 167} (1995)  301--350,
\href{http://arxiv.org/abs/arXiv:hep-th/9308122}{{\tt arXiv:hep-th/9308122}}.
%%CITATION = HEP-TH 9308122;%%.

\bibitem{MOsmall}
P.~S. Aspinwall, B.~R. Greene, and D.~R. Morrison, ``Measuring Small Distances
  In {$N{=}2$} Sigma Models,'' {\em Nucl. Phys. B} {\bf 420} (1994)  184--242,
\href{http://arxiv.org/abs/arXiv:hep-th/9311042}{{\tt arXiv:hep-th/9311042}}.
%%CITATION = HEP-TH 9311042;%%.

\bibitem{MOlooking}
D.~R. Morrison, ``Through the Looking Glass,'' in {\em Mirror Symmetry {III}},
  D.~H. Phong, L.~Vinet, and S.-T. Yau, eds., vol.~10 of {\em AMS/IP Stud. Adv.
  Math.}, pp.~263--277.
\newblock International Press, Cambridge, 1999.
\newblock \href{http://arxiv.org/abs/arXiv:alg-geom/9705028}{{\tt
  arXiv:alg-geom/9705028}}.

\bibitem{MOhhp}
M.~Herbst, K.~Hori, and D.~Page, ``Phases Of {$N=2$} Theories In {1+1}
  Dimensions With Boundary,''
\href{http://arxiv.org/abs/arXiv:0803.2045 [hep-th]}{{\tt arXiv:0803.2045
  [hep-th]}}.
%%CITATION = ARXIV:0803.2045;%%.

\bibitem{MOHori:2013ika}
K.~Hori and M.~Romo, ``Exact Results In Two-Dimensional (2,2) Supersymmetric
  Gauge Theories With Boundary,''
\href{http://arxiv.org/abs/1308.2438}{{\tt arXiv:1308.2438 [hep-th]}}.
%%CITATION = ARXIV:1308.2438;%%.

\bibitem{MOBatyrev1993}
V.~V. Batyrev, ``Dual Polyhedra and Mirror Symmetry for {C}alabi--{Y}au
  Hypersurfaces in Toric Varieties,'' {\em J. Algebraic Geom.} {\bf 3} (1994)
  no.~3, 493--535, \href{http://arxiv.org/abs/arXiv:alg-geom/9310003}{{\tt
  arXiv:alg-geom/9310003}}.

\bibitem{MOHowe:1986ys}
P.~S. Howe, G.~Papadopoulos, and K.~Stelle, ``Quantizing the {N=2} Super Sigma
  Model in Two-Dimensions,''
{\em Phys. Lett. B} {\bf 174} (1986)  405--410.
%%CITATION = PHLTA,B174,405;%%.

\bibitem{MOGross:1986iv}
D.~J. Gross and E.~Witten, ``Superstring Modifications of {E}instein's
  Equations,''
{\em Nucl. Phys. B} {\bf 277} (1986) no.~1, 1--10.
%%CITATION = NUPHA,B277,1;%%.

\bibitem{MOGrisaru:1986px}
M.~T. Grisaru, A.~E.~M. van~de Ven, and D.~Zanon, ``Four-Loop
  {$\beta$}-Function for the {$N=1$} and {$N=2$} Supersymmetric Nonlinear Sigma
  Model in Two Dimensions,''
{\em Phys. Lett. B} {\bf 173} (1986) no.~4, 423--428.
%%CITATION = PHLTA,B173,423;%%.

\bibitem{MOGrisaru:1986wj}
M.~T. Grisaru, D.~I. Kazakov, and D.~Zanon, ``Five-Loop Divergences for the
  {$N=2$} Supersymmetric Nonlinear Sigma-Model,''
{\em Nucl. Phys. B} {\bf 287} (1987) no.~1, 189--204.
%%CITATION = NUPHA,B287,189;%%.

\bibitem{MOZanon:1986gg}
D.~Zanon, ``Four-Loop {$\sigma$}-Model Beta-Functions Versus ${\alpha'}^3$
  Corrections to Superstring Effective Actions,'' in {\em Super Field Theories
  (NATO ASI Series)}, H.~C. Lee, V.~Elias, G.~Kunstatter, R.~B. Mann, and K.~S.
  Viswanathan, eds., vol.~160 of {\em Series B: Physics}, pp.~275--282.
\newblock Plenum Press,
1986.
\newblock
%%CITATION = HUTP-86/A054 ETC.;%%.

\bibitem{MOFreeman:1986zh}
M.~D. Freeman, C.~N. Pope, M.~F. Sohnius, and K.~S. Stelle, ``Higher-Order
  {$\sigma$}-Model Counterterms and the Effective Action for Superstrings,''
{\em Phys. Lett. B} {\bf 178} (1986) no.~2-3, 199--204.
%%CITATION = PHLTA,B178,199;%%.

\bibitem{MOBroadhurst:1996ur}
D.~J. Broadhurst, J.~Gracey, and D.~Kreimer, ``Beyond the Triangle and
  Uniqueness Relations: {N}onzeta Counterterms at Large {N} from Positive
  Knots,'' {\em Z. Phys. C} {\bf 75} (1997)  559--574,
\href{http://arxiv.org/abs/hep-th/9607174}{{\tt arXiv:hep-th/9607174
  [hep-th]}}.
%%CITATION = HEP-TH/9607174;%%.

\bibitem{MOMR1341859}
D.~Zagier, ``Values of Zeta Functions and Their Applications,'' in {\em First
  {E}uropean {C}ongress of {M}athematics, {V}ol.\ {II} ({P}aris, 1992)},
  vol.~120 of {\em Progr. Math.}, pp.~497--512.
\newblock Birkh{\"a}user, Basel, 1994.

\bibitem{MOMR2578167}
J.~Bl{\"u}mlein, D.~J. Broadhurst, and J.~A.~M. Vermaseren, ``The Multiple Zeta
  Value Data Mine,'' {\em Comput. Phys. Comm.} {\bf 181} (2010) no.~3,
  582--625, \href{http://arxiv.org/abs/arXiv:0907.2557 [math-ph]}{{\tt
  arXiv:0907.2557 [math-ph]}}.

\bibitem{MOarXiv:math.AG/9803119}
A.~Libgober, ``Chern Classes and the Periods of Mirrors,'' {\em Math. Res.
  Lett.} {\bf 6} (1999) no.~2, 141--149,
  \href{http://arxiv.org/abs/arXiv:math.AG/9803119}{{\tt
  arXiv:math.AG/9803119}}.

\bibitem{MOarXiv:0712.2204}
H.~Iritani, ``Real and Integral Structures in Quantum Cohomology {I}: {T}oric
  Orbifolds,'' \href{http://arxiv.org/abs/arXiv:0712.2204 [math.AG]}{{\tt
  arXiv:0712.2204 [math.AG]}}.

\bibitem{MOMR2553377}
H.~Iritani, ``An Integral Structure in Quantum Cohomology and Mirror Symmetry
  for Toric Orbifolds,'' {\em Adv. Math.} {\bf 222} (2009) no.~3, 1016--1079,
  \href{http://arxiv.org/abs/arXiv:0903.1463 [math.AG]}{{\tt arXiv:0903.1463
  [math.AG]}}.

\bibitem{MOarXiv:1101.4512}
H.~Iritani, ``Quantum Cohomology and Periods,'' {\em Ann. Inst. Fourier
  (Grenoble)} {\bf 61} (2011) no.~7, 2909--2958,
  \href{http://arxiv.org/abs/arXiv:1101.4512 [math.AG]}{{\tt arXiv:1101.4512
  [math.AG]}}.

\bibitem{MOMR2483750}
L.~Katzarkov, M.~Kontsevich, and T.~Pantev, ``Hodge Theoretic Aspects of Mirror
  Symmetry,'' in {\em From {H}odge Theory to Integrability and {TQFT}
  tt*-Geometry}, vol.~78 of {\em Proc. Sympos. Pure Math.}, pp.~87--174.
\newblock Amer. Math. Soc., Providence, RI, 2008.
\newblock \href{http://arxiv.org/abs/arXiv:0806.0107 [math.AG]}{{\tt
  arXiv:0806.0107 [math.AG]}}.

\bibitem{MOarXiv:1302.3760}
Y.~Honma and M.~Manabe, ``Exact {K}{\"a}hler Potential for {C}alabi--{Y}au
  Fourfolds,'' {\em JHEP} {\bf 1305} (2013)  102,
\href{http://arxiv.org/abs/1302.3760}{{\tt arXiv:1302.3760 [hep-th]}}.
%%CITATION = ARXIV:1302.3760;%%.

\bibitem{MOMR1397274}
B.~Dubrovin, ``Geometry of {$2$}{D} Topological Field Theories,'' in {\em
  Integrable Systems and Quantum Groups ({M}ontecatini {T}erme, 1993)},
  vol.~1620 of {\em Lecture Notes in Math.}, pp.~120--348.
\newblock Springer, Berlin, 1996.
\newblock
\href{http://arxiv.org/abs/arXiv:hep-th/9407018}{{\tt arXiv:hep-th/9407018}}.
\newblock
%%CITATION = HEP-TH/9407018;%%.

\bibitem{MOWitten:1991zz}
E.~Witten, ``Mirror Manifolds and Topological Field Theory,'' in {\em Essays on
  Mirror Manifolds}, pp.~120--158.
\newblock Int. Press, Hong Kong, 1992.
\newblock
\href{http://arxiv.org/abs/arXiv:hep-th/9112056}{{\tt arXiv:hep-th/9112056}}.
\newblock
%%CITATION = HEP-TH 9112056;%%.

\bibitem{MOtopgrav}
E.~Witten, ``On the Structure of the Topological Phase of Two-Dimensional
  Gravity,'' {\em Nucl. Phys. B} {\bf 340} (1990)  281--332.

\bibitem{MOVafa:1991uz}
C.~Vafa, ``Topological Mirrors and Quantum Rings,'' in {\em Essays on Mirror
  Manifolds}, pp.~96--119.
\newblock Int. Press, Hong Kong, 1992.
\newblock
\href{http://arxiv.org/abs/arXiv:hep-th/9111017}{{\tt arXiv:hep-th/9111017}}.
\newblock
%%CITATION = HEP-TH/9111017;%%.

\bibitem{MOMR1144529}
E.~Witten, ``Two-Dimensional Gravity and Intersection Theory on Moduli Space,''
  in {\em Surveys in Differential Geometry (Cambridge, MA, 1990)},
  pp.~243--310.
\newblock Lehigh Univ., Bethlehem, PA,
1991.
\newblock
%%CITATION = 00078,1,243;%%.

\bibitem{MODijkgraaf:1990dj}
R.~Dijkgraaf, H.~L. Verlinde, and E.~P. Verlinde, ``Topological Strings in {$d
  < 1$},''
{\em Nucl. Phys. B} {\bf 352} (1991)  59--86.
%%CITATION = NUPHA,B352,59;%%.

\bibitem{MODubrovin:1992dz}
B.~Dubrovin, ``Integrable Systems in Topological Field Theory,''
{\em Nucl. Phys. B} {\bf 379} (1992)  627--689.
%%CITATION = NUPHA,B379,627;%%.

\bibitem{MOhep-th/9402147}
M.~Kontsevich and Y.~Manin, ``{G}romov-{W}itten classes, Quantum Cohomology,
  and Enumerative Geometry,'' {\em Commun. Math. Phys.} {\bf 164} (1994)
  525--562, \href{http://arxiv.org/abs/arXiv:hep-th/9402147}{{\tt
  arXiv:hep-th/9402147}}.

\bibitem{MORuanTian}
Y.~Ruan and G.~Tian, ``A Mathematical Theory of Quantum Cohomology,'' {\em J.
  Differential Geom.} {\bf 42} (1995) no.~2, 259--367.

\bibitem{MOMR1467172}
J.~Li and G.~Tian, ``Virtual Moduli Cycles and {G}romov-{W}itten Invariants of
  Algebraic Varieties,'' {\em J. Amer. Math. Soc.} {\bf 11} (1998) no.~1,
  119--174, \href{http://arxiv.org/abs/arXiv:alg-geom/9602007}{{\tt
  arXiv:alg-geom/9602007}}.

\bibitem{MOtftrc}
P.~S. Aspinwall and D.~R. Morrison, ``Topological Field Theory and Rational
  Curves,'' {\em Commun. Math. Phys.} {\bf 151} (1993)  245--262,
\href{http://arxiv.org/abs/arXiv:hep-th/9110048}{{\tt arXiv:hep-th/9110048}}.
%%CITATION = HEP-TH 9110048;%%.

\bibitem{MOhigherD}
B.~R. Greene, D.~R. Morrison, and M.~R. Plesser, ``Mirror Manifolds in Higher
  Dimension,'' {\em Commun. Math. Phys.} {\bf 173} (1995)  559--598,
  \href{http://arxiv.org/abs/arXiv:hep-th/9402119}{{\tt arXiv:hep-th/9402119}}.

\bibitem{MOKlemm:2007in}
A.~Klemm and R.~Pandharipande, ``Enumerative Geometry of {C}alabi-{Y}au
  4-Folds,'' {\em Commun. Math. Phys.} {\bf 281} (2008)  621--653,
\href{http://arxiv.org/abs/arXiv:math.AG/0702189}{{\tt arXiv:math.AG/0702189}}.
%%CITATION = MATH/0702189;%%.

\bibitem{MOGoddard:1976qe}
P.~Goddard, J.~Nuyts, and D.~I. Olive, ``Gauge Theories and Magnetic Charge,''
{\em Nucl. Phys. B} {\bf 125} (1977)  1--28.
%%CITATION = NUPHA,B125,1;%%.

\bibitem{MOHonda:2013uca}
D.~Honda and T.~Okuda, ``Exact Results for Boundaries and Domain Walls in 2d
  Supersymmetric Theories,'' {\em JHEP} {\bf 09} (2015)  140,
\href{http://arxiv.org/abs/1308.2217}{{\tt arXiv:1308.2217 [hep-th]}}.
%%CITATION = ARXIV:1308.2217;%%.

\bibitem{MOSugishita:2013jca}
S.~Sugishita and S.~Terashima, ``Exact Results in Supersymmetric Field Theories
  on Manifolds with Boundaries,'' {\em JHEP} {\bf 11} (2013)  021,
\href{http://arxiv.org/abs/1308.1973}{{\tt arXiv:1308.1973 [hep-th]}}.
%%CITATION = ARXIV:1308.1973;%%.

\bibitem{MOKim:2013ola}
H.~Kim, S.~Lee, and P.~Yi, ``Exact Partition Functions on {$\mathbb{RP}^2$} and
  Orientifolds,'' {\em JHEP} {\bf 02} (2014)  103,
\href{http://arxiv.org/abs/1310.4505}{{\tt arXiv:1310.4505 [hep-th]}}.
%%CITATION = ARXIV:1310.4505;%%.

\bibitem{MOBecker:1995kb}
K.~Becker, M.~Becker, and A.~Strominger, ``Fivebranes, Membranes and
  Nonperturbative String Theory,'' {\em Nucl. Phys. B} {\bf 456} (1995)
  130--152,
\href{http://arxiv.org/abs/hep-th/9507158}{{\tt arXiv:hep-th/9507158}}.
%%CITATION = HEP-TH 9507158;%%.

\bibitem{MOOoguri:1996ck}
H.~Ooguri, Y.~Oz, and Z.~Yin, ``D-Branes on {C}alabi--{Y}au Spaces and their
  Mirrors,'' {\em Nucl. Phys. B} {\bf 477} (1996)  407--430,
\href{http://arxiv.org/abs/arXiv:hep-th/9606112}{{\tt arXiv:hep-th/9606112}}.
%%CITATION = HEP-TH 9606112;%%.

\bibitem{MOKapustin:2009kz}
A.~Kapustin, B.~Willett, and I.~Yaakov, ``Exact Results for {W}ilson Loops in
  Superconformal {C}hern-{S}imons Theories with Matter,'' {\em JHEP} {\bf 1003}
  (2010)  089,
\href{http://arxiv.org/abs/0909.4559}{{\tt arXiv:0909.4559 [hep-th]}}.
%%CITATION = ARXIV:0909.4559;%%.

\bibitem{MOHama:2011ea}
N.~Hama, K.~Hosomichi, and S.~Lee, ``{SUSY} Gauge Theories on Squashed
  Three-Spheres,'' {\em JHEP} {\bf 1105} (2011)  014,
\href{http://arxiv.org/abs/1102.4716}{{\tt arXiv:1102.4716 [hep-th]}}.
%%CITATION = ARXIV:1102.4716;%%.

\bibitem{MOPasquetti:2011fj}
S.~Pasquetti, ``Factorisation of {N = 2} Theories on the Squashed 3-Sphere,''
  {\em JHEP} {\bf 1204} (2012)  120,
\href{http://arxiv.org/abs/1111.6905}{{\tt arXiv:1111.6905 [hep-th]}}.
%%CITATION = ARXIV:1111.6905;%%.

\bibitem{MOBeem:2012mb}
C.~Beem, T.~Dimofte, and S.~Pasquetti, ``Holomorphic Blocks in Three
  Dimensions,'' {\em JHEP} {\bf 1412} (2014)  177,
\href{http://arxiv.org/abs/1211.1986}{{\tt arXiv:1211.1986 [hep-th]}}.
%%CITATION = ARXIV:1211.1986;%%.

\bibitem{MOCecotti:1991me}
S.~Cecotti and C.~Vafa, ``Topological Antitopological Fusion,''
{\em Nucl. Phys. B} {\bf 367} (1991)  359--461.
%%CITATION = NUPHA,B367,359;%%.

\bibitem{ContributionBL}
F.~Benini and B.~{Le Floch}, ``Supersymmetric localization in two dimensions,''
  {\em Journal of Physics A} {\bf xx} (2016)  000,
  \href{http://arxiv.org/abs/1608.02955}{{\tt 1608.02955}}.

\bibitem{MOGH}
P.~Griffiths and J.~Harris, {\em Principles of Algebraic Geometry}.
\newblock John Wiley \& Sons, New York, 1978.

\bibitem{MOtsikh}
A.~K. Tsikh, {\em Multidimensional Residues and their Applications}, vol.~103
  of {\em Translations of Mathematical Monographs}.
\newblock Amer. Math. Soc., Providence, R. I., 1992.

\bibitem{MOPassare:1996db}
M.~Passare, A.~K. Tsikh, and A.~A. Cheshel, ``Multiple {M}ellin--{B}arnes
  Integrals as Periods of {C}alabi--{Y}au Manifolds with Several Moduli,'' {\em
  Theor. Math. Phys.} {\bf 109} (1997)  1544--1555,
\href{http://arxiv.org/abs/arXiv:hep-th/9609215}{{\tt arXiv:hep-th/9609215}}.
%%CITATION = HEP-TH 9609215;%%.

\bibitem{MOmath.AG/9912109}
R.~P. Horja, {\em Hypergeometric Functions and Mirror Symmetry in Toric
  Varieties}.
\newblock PhD thesis, Duke University, 1999.
\newblock \href{http://arxiv.org/abs/arXiv:math.AG/9912109}{{\tt
  arXiv:math.AG/9912109}}.

\bibitem{MOMR2271990}
L.~A. Borisov and R.~P. Horja, ``Mellin-{B}arnes Integrals as {F}ourier-{M}ukai
  Transforms,'' {\em Adv. Math.} {\bf 207} (2006) no.~2, 876--927,
  \href{http://arxiv.org/abs/arXiv:math.AG/0510486}{{\tt
  arXiv:math.AG/0510486}}.

\bibitem{MOBatyrev-Borisov-1994}
V.~V. Batyrev and L.~A. Borisov, ``On {C}alabi--{Y}au Complete Intersections in
  Toric Varieties,'' in {\em Higher-Dimensional Complex Varieties (Trento,
  1994)}, pp.~39--65.
\newblock de Gruyter, Berlin, 1996.
\newblock \href{http://arxiv.org/abs/arXiv:alg-geom/9412017}{{\tt
  arXiv:alg-geom/9412017}}.

\bibitem{MOclemens1983double}
C.~H. Clemens, ``Double Solids,'' {\em Adv. in Math.} {\bf 47} (1983) no.~2,
  107--230.

\bibitem{MOMR848512}
R.~Friedman, ``Simultaneous Resolution of Threefold Double Points,'' {\em Math.
  Ann.} {\bf 274} (1986) no.~4, 671--689.

\bibitem{MOMR909231}
M.~Reid, ``The Moduli Space of {$3$}-Folds with {$K=0$} May Nevertheless Be
  Irreducible,'' {\em Math. Ann.} {\bf 278} (1987) no.~1-4, 329--334.

\bibitem{MOCandelas:1989ug}
P.~Candelas, P.~S. Green, and T.~Hubsch, ``Rolling Among {C}alabi--{Y}au
  Vacua,''
{\em Nucl. Phys. B} {\bf 330} (1990)  49--102.
%%CITATION = NUPHA,B330,49;%%.

\bibitem{MObhole}
B.~R. Greene, D.~R. Morrison, and A.~Strominger, ``Black Hole Condensation and
  the Unification of String Vacua,'' {\em Nucl. Phys. B} {\bf 451} (1995)
  109--120,
\href{http://arxiv.org/abs/arXiv:hep-th/9504145}{{\tt arXiv:hep-th/9504145}}.
%%CITATION = HEP-TH 9504145;%%.

\bibitem{MOMR1839289}
A.-M. Li and Y.~Ruan, ``Symplectic Surgery and {G}romov-{W}itten Invariants of
  {C}alabi-{Y}au 3-Folds,'' {\em Invent. Math.} {\bf 145} (2001) no.~1,
  151--218, \href{http://arxiv.org/abs/arXiv:math.AG/9803036}{{\tt
  arXiv:math.AG/9803036}}.

\bibitem{MOMR1201748}
A.~Libgober and J.~Teitelbaum, ``Lines on {C}alabi--{Y}au Complete
  Intersections, Mirror Symmetry, and {P}icard--{F}uchs Equations,'' {\em
  Internat. Math. Res. Notices} (1993) no.~1, 29--39,
  \href{http://arxiv.org/abs/arXiv:alg-geom/9301001}{{\tt
  arXiv:alg-geom/9301001}}.

\bibitem{MOMR701704}
J.~Wess and J.~Bagger, {\em Supersymmetry and Supergravity}.
\newblock Princeton University Press, Princeton, N.J., 1983.

\end{thebibliography}
